\documentclass[epj]{svjour}
\usepackage[latin2]{inputenc}
\usepackage[fleqn]{amsmath}
\usepackage{amssymb,amsfonts}
\usepackage{mathrsfs}
\usepackage{dsfont}
\usepackage{graphicx,bm}
\usepackage{multirow}
\usepackage{color}
\usepackage{t1enc}
\usepackage{enumerate}
\usepackage{widetext}
\usepackage{epstopdf}

\newcommand{\eref}[1]{Eq.~\eqref{#1}}
\newcommand{\figref}[1]{Fig.~\ref{#1}}
\newcommand{\secref}[1]{Sec.~\ref{#1}}
\newcommand{\apref}[1]{Appendix~\ref{#1}}
\newcommand{\tabref}[1]{Table~\ref{#1}}

\allowdisplaybreaks
\begin{document}

\title{A statistical method to estimate low-energy hadronic cross sections}

\author{G{\'a}bor Balassa\inst{1} \and P{\'e}ter Kov{\'a}cs\inst{1,2} \and Gy{\"o}rgy Wolf\inst{1}}                

\institute{Institute for Particle and Nuclear Physics, Wigner Research
  Centre for Physics, Hungarian Academy of Sciences, H-1525 Budapest,
  Hungary \and ExtreMe Matter Instititute EMMI, GSI Helmholtzzentrum
  f{\"u}r Schwerionenforschung, Planckstrasse 1, 64291 Darmstadt,
  Germany}

\date{Received: date / Revised version: date}

\abstract{In this article we propose a model based on the Statistical
  Bootstrap approach to estimate the cross sections of different
  hadronic reactions up to a few GeV in c.m.s energy. The method is
  based on the idea, when two particles collide a so called fireball
  is formed, which after a short time period decays statistically into
  a specific final state. To calculate the probabilities we use a
  phase space description extended with quark combinatorial factors
  and the possibility of more than one fireball formation. In a few
  simple cases the probability of a specific final state can be
  calculated analytically, where we show that the model is able to
  reproduce the ratios of the considered cross sections. We also show
  that the model is able to describe proton\,-\,antiproton annihilation at
  rest. In the latter case we used a numerical method to calculate the
  more complicated final state probabilities. Additionally, we
  examined the formation of strange and charmed mesons as well, where
  we used existing data to fit the relevant model parameters.
} 

\maketitle

\section{Introduction}
\label{Sec:intro}

Transport models are very important tools to understand the dynamics
of heavy ion collisions, plan the detectors, or even understand the
experimental results. Very essential ingredients of these models are
the elementary hadronic cross sections. Some of these are
experimentally well known in a broad energy range, however, cross
sections involving short lifetime particles cannot be determined
experimentally. Below $1-2$~GeV effective field theories can be used
\cite{Scherer:2002,Caswell:1985}, however, the many resonances --
which are not part of the model -- restrict the energy range of
validity of these methods. An interesting attempt was made by J.~Van
de Wiele and S.~Ong \cite{Wiele:2010} to use Regge propagators in
higher energies to extend the low energy effective field theories.

Statistical methods are useful tools to determine particle
multiplicities in high energy collisions. The first attempt can be
associated with Fermi (1950) who proposed a simple phase space method
to describe the cross sections of binary collisions
\cite{Fermi:1950}. Fermi estimated the probabilities of one and
multiple pion creation in nucleon\,-\,nucleon collisions based solely
on statistical considerations. Although, he treated the momentum and
angular momentum conservation only approximately, he could describe
some of the cross sections correctly.

Latter R. Hagedorn developed the statistical bootstrap method
\cite{Hagedorn:1965,Hagedorn:1968}, which was able to explain hadron
multiplicities. The basic assumption of Hagedorn's theory is that
asymptotically ($m \to \infty$, where $m$ is the invariant mass) no
particles are elementary, but instead every particle is made up of all
resonances and particles, that is at very high energies ``fireballs
(hot hadrons) are made of fireballs, are made of fireballs
$\dots$''. This condition leads to the so\,-\,called bootstrap
equation, which can be solved to get asymptotically the density of
states ($\rho(m)$) inside a fireball
\cite{Frautschi:1971,Nahm:1972,Yellin:1973}. Once $\rho(m)$ is known,
the partition function is readily given by a Laplace transformation,
$\mathcal{Z}(T)=\int dm \rho(m) e^{-m/T}$.  From the partition
function the average number of particles can be calculated, therefore
the particle multiplicities in high energy collisions can be obtained.

For sufficiently high energies statistical methods
\cite{Becattini:1997a,Becattini:1997b,Degtyarenko:2000,Becattini:2004if,Becattini:2004rq,Bignamini:2012,Becattini:2012zza} show good agreement with existing
data, however, at lower energies where interesting non - perturbative effects
can be important, and the particle multiplicities are low, the number of
results are somewhat less (for an exception see \cite{Ferroni:2011}, in which
the authors could reach a fairly good agreement with experimental data in the
$2.1$ to $2.6$~GeV energy range by using a one fireball model).
It is worth to note that this behavior can be caused by the fact that
at high energies (few hundred GeV) the quark masses become negligible
and certain quantities like the quark creation probabilities and quark
pair numbers -- see later -- become the same for each
flavor. Consequently, at such high energies the effects of new
ingredients introduced here could become negligible or cancel out from
the calculated cross section ratios.

In our case, we would like to calculate hadronic cross section of such
reactions that have a relatively high energy ($2-10$~GeV) and can have
more than two particles in the final state. It is worth to note that
the calculated elementary, experimentally unknown cross sections can
be used to improve our existing BUU
(Boltzmann\,-\,Uehling\,-\,Uhlenbeck) code for simulating
Heavy\,-\,ion collisions \cite{Wolf:1990_1993_1997}. Thus, we tried to
extend a statistical model to reach our goals. Our method incorporate
that during the collision a compound system, a so\,-\,called fireball,
is formed and after a short time, through possible production of
subsequent fireballs, this system decays into a specific final
state. In our approach the probability of the resulting final state
can be calculated from the corresponding phase space, the quark
content of the final state, the density of states $\rho(m)$, and from
the properties of the fireballs.

The paper is organized as follows. In Section \ref{Sec:basics} we
introduce the basic mathematical formulation of the model and in
Section 3 we show that it can be used to describe medium energy
reactions.  For that reason, a Monte Carlo algorithm was developed to
describe final states with more than 3 particles, in case of which
multiple fireball decays could be important. The method is used to
describe proton\,-\,antiproton annihilation at rest ,where a very good
agreement with the existing data is found. Exotic meson -- containing
strange and/or charm quarks -- formation is also examined, in case of
which the model parameters are fitted to the existing experimental
data. Finally, we conclude in Section 4.

\section{Basic formulation}
\label{Sec:basics}

The bootstrap idea is very useful in describing particle
multiplicities, however, if the probability of a specific reaction is
to be calculated, the basic method should be extended. For more on the
basics of the statistical method see for instance
\cite{Rafelski:2016nxx}. As it was already mentioned, our starting
point is the fireball formation and its subsequent decay statistically
into different final states. It is worth to summarize here the most
important ingredients of our approach, which amounts in the
multi\,-\,fireball decay scheme and the quark combinatorial
factors. Some version of multi\,-\,fireball creation was already used
{\it e.g.} in \cite{Sitte:2012,Bediaga:1990hz}, however, our decay
scheme is different, which will be discussed in detail below. On the
other hand, the quark combinatorial factors, introduced in the last
part of this section -- as far as we are aware -- has not been
used together with the bootstrap idea.

Our basic assumption is that the generalized cross section
($\sigma^{n\to k}(M)$) of the $n\rightarrow k$ reaction -- containing
$n$ incoming and $k$ outgoing particles -- can be factorized into two
terms, one which describes the dynamics of the collision, and the
other one with mixed dynamical and statistical part, which describes
the probabilities of each final state channel,
\begin{equation}
  \begin{split}
    \label{Eq:sigma_factor}
    \sigma^{n\to k}(M)=&\left(\int \prod_{i=1}^{n} d^3p_i R(M,p_1,\dots,p_n)\right) \\
    &\quad\quad\quad\times \left(\int \prod_{j=1}^{k} d^3q_j w(M,q_1,\dots,q_k)\right),
  \end{split}
\end{equation}
where $M=\sqrt{s}$ is the CM energy of the colliding particles, while
$p_i$ and $q_i$ stand for the incoming and outgoing momenta,
respectively. As a further simplification we assume that the dynamical
term can be substituted by the integrated total cross section of the
considered reaction. Moreover, for the mixed term it is assumed that
the energy\,-\,momentum conservation can be factored out, that is
  \begin{eqnarray}
    \negthickspace\int \prod_{i=1}^{n} d^3p_i R(M,p_1,\dots,p_n) &\approx&
    \sigma^{(n)}_{\text{Tot}},\label{Eq:R_approx} \\ 
    \negthickspace\int \prod_{j=1}^{k} d^3q_j w(M,q_1,\dots,q_k) &\approx& \tilde
    w(M) \frac{\Phi_k}{V^{k-1}}\equiv W(M), \label{Eq:W_approx}
  \end{eqnarray}
where
\begin{equation}
  \label{Eq:ph_space}
  \begin{split}
    \Phi_k(M,m_1,\dots,m_k) =& V^{k-1}\\
    \times\int \prod_{i=1}^{k} d^3q_i\delta
    &\left(\sum_{j=1}^{k}E_j-M\right)
    \delta\left(\sum_{l=1}^{k}\mathbf{q}_l\right)
  \end{split}
\end{equation}
is the $k$ particle phase space, $\sigma^{(n)}_{\text{Tot}}$ is the
generalized total cross section of $n$ incoming particles and $V$ is
the interaction volume, which is set to $V=4\pi/(3m_{\pi}^{3})$. It is
worth to note that if we want to give e.g. the angular dependence of
the outgoing particles we assume every possibility may happen with a
probability weighted by the corresponding phase space. This assumption
seems to be a very crude one, but it is the main idea behind other
statistical methods in the literature (see e.g. \cite{Batko:1992}) and
we will show in \secref{Subsec:final_ud} that despite its simplicity
it can give reasonable results. The model can be extended to determine
not only total, but also generalized differential cross sections,
however, this extension is not necessary to describe the basics of our
model. In the current general form of
Eqs.~\eqref{Eq:sigma_factor}\,-\,\eqref{Eq:ph_space} the model could
be used to describe many body collisions, which we plan to do in a
subsequent work, however, we now only concentrate on two body
collisions. Consequently, from now on \eref{Eq:R_approx} has two
incoming momenta $p_1$, and $p_2$, while
Eqs.~\eqref{Eq:W_approx}\,-\,\eqref{Eq:ph_space} hold. Based on
Eqs.~\eqref{Eq:sigma_factor}\,-\,\eqref{Eq:ph_space} we calculate the
considered partial cross section according to
\begin{equation}
\sigma(M) = W(M)\cdot\sigma_{Tot}(M),
\end{equation}
where $\sigma(M)(\equiv \sigma^{2\to k}(M))$ and
$\sigma_{\text{Tot}}(\equiv \sigma^{(2)}_{\text{Tot}})$ are the usual
partial and total cross sections respectively. The later is usually
taken from experimental data while $W(M)$, which is the total
formation probability of the desired end state, is calculated from our
model. More precisely, instead of calculating $W(M)$'s directly, we
always use some reference channel, with which we only calculate the
ratios of $W(M)$'s. Let us demonstrate this in the following
example. Consider the reaction $p\pi^- \rightarrow n\pi^+ \pi^-$. To
calculate its cross section, we use the reference channel $p\pi^-
\rightarrow p\pi^-$. It is assumed that this elastic cross section is
known from somewhere else. Assuming our factorization scheme works,
and we know the cross section of the reference channel, we can
calculate the desired cross section as follows
\begin{multline}
  \sigma_{p\pi^{-} \rightarrow n\pi^+ \pi^{-}} \equiv 
  \frac{\sigma_{p\pi^{-}\to n\pi^{+}\pi^{-}}}{\sigma_{p\pi^{-}\to
      p\pi^{-}}}\sigma_{p\pi^{-} \to p\pi^{-}} \\
  = \frac{W_{n\pi^{+}\pi^{-}}}{W_{p\pi^{-}}}
  \frac{\sigma_{p\pi^{-} }^{Tot}}{\sigma_{p\pi^{-}}^{Tot}} \sigma_{p\pi^{-}\to
    p\pi^{-}} \\
  = \frac{W_{n\pi^{+}\pi^{-}}}{W_{p\pi^{-}}}\sigma_{p\pi^{-}\to
    p\pi^{-}},
\end{multline}
here the only unknown part is the ratio
$\frac{W_{n\pi^{+}\pi^{-}}}{W_{p\pi^{-}}}$, which can be calculated
much easier then the $W$'s themselves, since common factors -- for
instance in the case of the same initial states the rather complicated
normalization factor -- cancels out. The remaining part of the section
will be dedicated to the explanation of how we calculate the $W$'s.

It is worth to note that in the literature there are various
approaches like cases where authors considered only one, see {\it
  e.g.}  \cite{Johns:1972bi}, or more than one, see {\it e.g.}
\cite{Sitte:2012,Bediaga:1990hz}, fireball(s) formation. In our
approach we allow more than one fireball to form in a chain of
consecutive decays, that is in every step -- until the energy runs out
-- the fireball splits into two. This is a similar approach as in the
model of Frautschi \cite{Frautschi:1971}.  Generally, when $k$
fireball is formed after the collision with $M$ invariant mass, and
from the individual fireballs $n_1,n_2,\dots,n_k$ hadrons are
produced, the total formation probability $W_k^{n_1,\dots,n_k}$ can be
written as,

\begin{widetext}
  \begin{equation}
    \label{Eq:W_calc}
  W_k^{n_1,\dots,n_k}(M) = \mathcal{N}_k(M) P^{\text{fb}}_{k}(M) C_Q(M)
  \int\limits_{x_{1,min}}^{x_{1,max}}\cdots\int\limits_{x_{k,min}}^{x_{k,max}}
  \prod_{i=1}^{k}dx_i P^{H,1}_{n_1}(x_1)P^{H,2}_{n_2}(x_2)\cdots
  P^{H,k}_{n_k}(x_k)\delta\left(\sum_{i=1}^{k}x_i - M\right),
\end{equation}
\end{widetext}

where $\mathcal{N}_k(M)$ is a normalization factor\footnote{It should
  be noted that the normalization factor depends on the initial
  state.}, $P^{\text{fb}}_{k}$ stands for the formation probability of
$k$ fireballs, $C_Q(M)$ is the quark\,-\,combinatorial factor, $x
_i$'s are the invariant masses of the individual fireballs,
$P^{H,i}_{n_i}(x_i)$'s are the hadronization probabilities of the
$i$th fireball that is producing $n_i$ hadrons, while $x_{i,min}$ and
$x_{i,max}$ are the lower and upper kinematical limits of the decay of
the $i$th fireball, which are imposed by the choice of the final
states. $P^{\text{fb}}_{k}$, $P^{H,i}_{n_i}$, and $C_Q$ will be
discussed below in detail.  We assume for the hadronization that only
two and three body decays may take place following the reasoning of
Frautschi \cite{Frautschi:1971}. He calculated that the probability of
a fireball decay into $n$ hadrons -- in contrast to Hagedorn's picture
\cite{Hagedorn:1965,Hagedorn:1968} -- is energy independent and the
dominant ones are the $n=2, 3$ body decays with $P^d_2=0.69$ and
$P^d_3=0.24$ probabilities. It is worth to note that we assume --
based on \cite{Frautschi:1971} -- that the minimal number of hadrons
coming out of a fireball is $2$. According to that the $n_i$'s in
\eref{Eq:W_calc} can take only the values $2$ or $3$, while the
hadronization probabilities $P^{H,i}_{2}(x_i)$ and $P^{H,i}_{3}(x_i)$
can be written as
\begin{eqnarray}
  P^{H,i}_{2}(x_i) &=& \prod_{l=1}^2(2s_l+1) P^d_2
  \frac{\Phi_2(x_i,m_1,m_2)}{\rho(x_i)(2\pi)^3
    N_I!}, \label{Eq:PH2}\\
  P^{H,i}_{3}(x_i) &=& \prod_{l=1}^3(2s_l+1)
  P^d_3 \frac{\Phi_3(x_i,m_1,m_2,m_3)}{\rho(x_i)(2\pi)^6 N_I!},\label{Eq:PH3}
\end{eqnarray}
where $s_l$ and $m_l$ are the spin and mass of the $l^{\text{th}}$
outgoing physical particle, respectively, $N_I$ is the number of
identical particles in the final state, while $\rho(x_i)$ stands for
the density of states with invariant mass $x_i$. The explicit form of
$\rho(M)$ is taken from \cite{Hagedorn:1973} and it reads
\begin{equation}
  \rho(M)=\frac{a\sqrt{M}}{(M_0+M)^{3.5}}  e^{\frac{M}{T_0}},
\end{equation}
where $a, M_0$, and $T_0$ are free parameters. In our calculations,
however, only $M_0$ and $T_0$ are relevant, since $a$ always cancels
out from the ratio of the $W$'s. Their values are given by
$M_0=500$~MeV and $T_0=130-170$~MeV \cite{Hagedorn:1973}. For the
latter, we calculated the
  \begin{equation}
    \begin{split}
      \frac{\sigma_{p\bar p\to n\bar n}}{\sigma_{p\bar p\to \pi^+\pi^-}},
      \frac{\sigma_{p\bar p\to p\bar p\pi^0}}{\sigma_{p\bar p\to\pi^+\pi^-}}&,
      \frac{\sigma_{p\bar p\to\pi^+\pi^-}}{\sigma_{p\bar p\to\Lambda\bar\Lambda}},\\
      &\frac{\sigma_{p\bar p\to\pi^+\pi^-}}{\sigma_{p \bar p\to K^+ K^-}},\,
      \text{and}\; \frac{\sigma_{p p\to p p\rho^0}}{\sigma_{p p\to n \Delta^{++}}}\nonumber
    \end{split}
  \end{equation}
ratios and varied $T_0$ in order to get a good agreement with the
experimentally known values, which lead finally to $T_0=160$~MeV.

\begin{figure*}[!t] 
  \centerline{\includegraphics[width=0.65\textwidth]{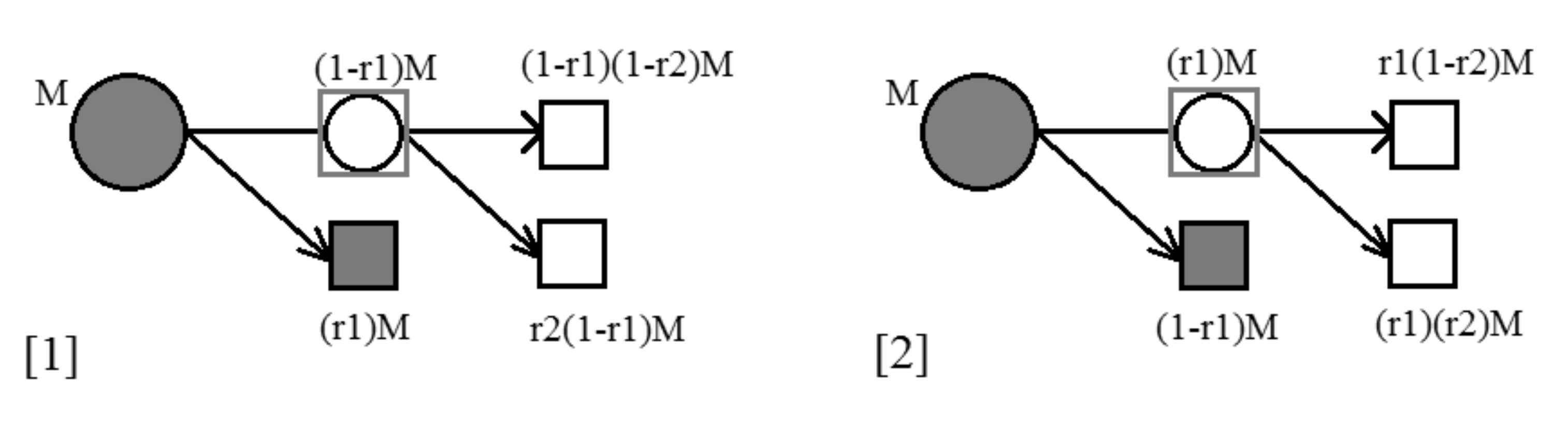}}
  \caption{The decay scheme for the two fireballs formation case. Left
    and right figure shows the two possible subcases.}
  \label{Fig:2fireball}     
\end{figure*}

\begin{figure}[!t]
  \includegraphics[width=0.45\textwidth]{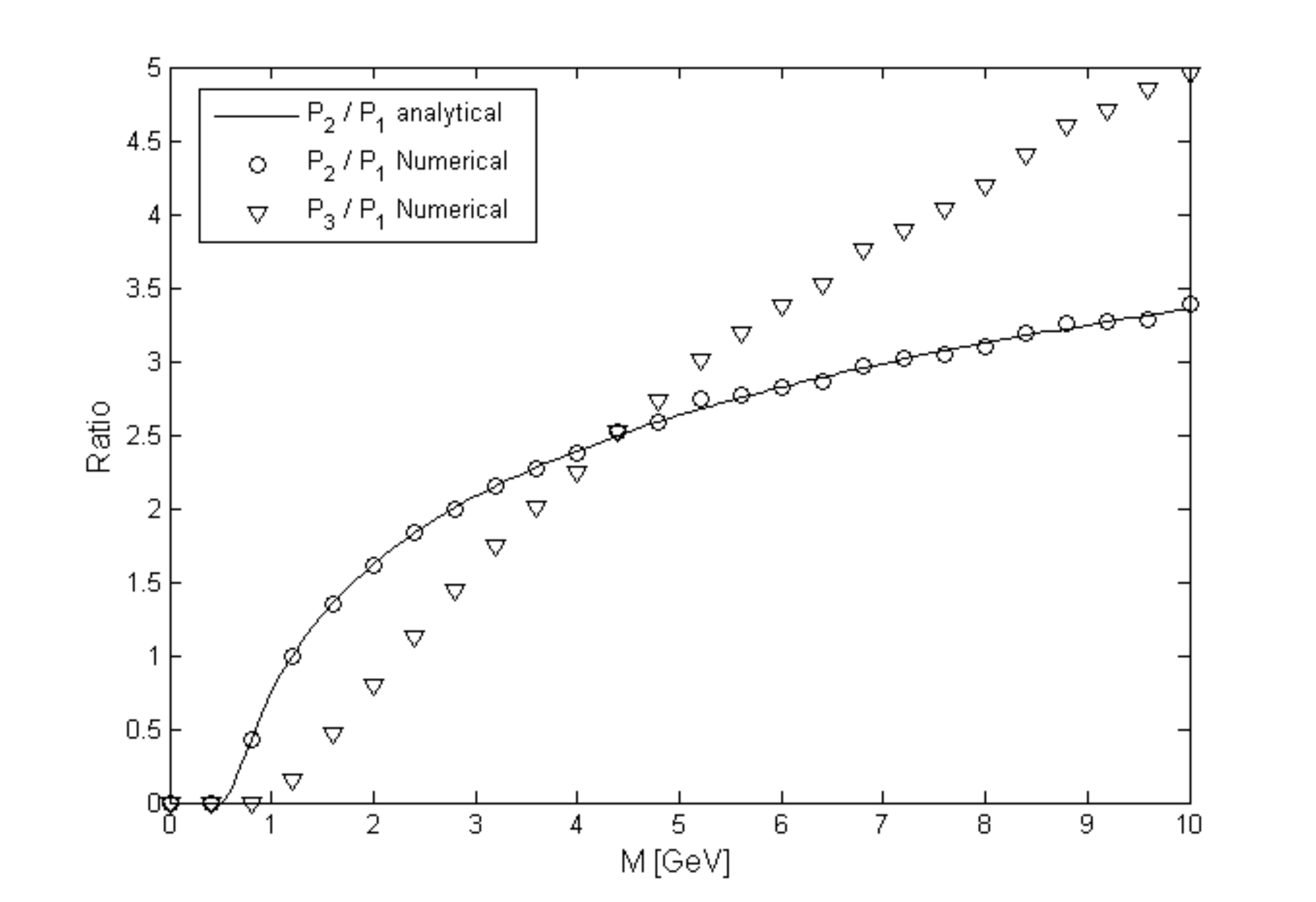}
\caption{Ratio of the fireball formation probabilities
  $P^{\text{fb}}_{2}/P^{\text{fb}}_{1}$ and
  $P^{\text{fb}}_{3}/P^{\text{fb}}_{1}$ as function of the invariant
  mass. Analytic and Monte\,-\,Carlo simulation results are shown by
  solid line and circles/triangles, respectively.}
\label{Fig:P2overP1}
\end{figure}

Let's turn to the fireball formation probability
$P^{\text{fb}}_{k}$. Either we can consider these probabilities as
energy dependent free parameters of the model, which we can fit from
measured data, or we can calculate them based on a purely statistical
approach. Thus, we calculated the probabilities of one two and three
fireballs formation, since four and more fireballs formation
  would give significant contribution only at larger energies.
  \footnote{In some cases, like $p-\bar p$ annihilation, where
    multiple particles creation ($\ge 10$) is possible even at lower
    energies it can happen that four fireballs formation becomes
    significant.} It is important to note here that, however,
  while in the case of analytic calculations we restricted our
  calculations only to maximum three fireballs, we also used a
  Monte\,-\,Carlo simulation, where this restriction was abolished. According
  to the simulations, we can justify that for the currently considered
  processes and energy range the inclusion of four or more fireballs
  have a negligible impact on the results. Here we assumed a chain
  like decay scheme -- in one step one fireball can split into two
  smaller. The method is shown for the two fireballs formation case in
  \figref{Fig:2fireball}, where two different subcases are possible.
In the figure the circle marked with $M$ is the initial fireball,
which tries to decay into more fireballs. Here $M$ denotes the invariant mass,
which is the available energy from the collision. Next we assume that
this fireball can decay into two more fireballs but one of them is
permanent -- denoted by a filled square in \figref{Fig:2fireball} --,
which means it cannot decay any further. However, the other fireball
can decay into another two fireballs, and so on. Those fireballs that
can not decay any further stay and subsequently hadronize giving a
final state with the previously described probabilities. Each
permanent fireball will give some particles to the final state, so the
net final state will be the sum of those particles coming from the
different permanent fireballs. Back to our example case, we suppose
that the permanent fireball is produced with invariant mass $(r_1M)$
(left panel of \figref{Fig:2fireball}), where $(r_1\in U[0,1])$ is a
random number with uniform distribution. This fireball will hadronize,
consequently it must have a minimum invariant mass that can give a
real final state, which will be denoted by $m_c$. Practically, since
minimum two particles will be produced from a fireball, we can assume
that $m_c = 2m_{\pi}$. The maximum is also limited, because the
remaining invariant mass also need to produce at least two particles.
The other fireball with invariant mass $((1-r_1)M)$ in principle tries
to decay further, but in this case it can not, because at the end
exactly two permanent fireballs are needed. Thus, we have to make sure
that this fireball is unable decay any further, but it also has to
produce at least two particles, which constrain both $r_1$ and $r_2$. The
unfilled squares in \figref{Fig:2fireball} means that those fireballs
cannot be created. A fireball can not decay any further if either one
of its subsequent fireballs fails to have at least $m_c$ invariant
mass (practically can not produce two pions). With this in mind the
constraints for the first case are the following
\begin{eqnarray}
  r_1M > m_{c},&\\
  (1-r_1)M > m_{c},&\\
  (1-r_1)(1-r_2)M < m_{c}, \;\text{or}\; &(1-r_1) r_2M < m_{c}.
\end{eqnarray}
The same reasoning is valid for the right panel in
\figref{Fig:2fireball} where the only difference is that in the first
step the fireballs with invariant masses $r_1M$ and $(1-r_1)M$ should
be swapped. The probability for this decay scheme can be calculated
easily if we define the following five events ($A_0-A_5$),
\begin{eqnarray}
  &A_0:\left(r_1>\frac{m_{c}}{M} \right) \;\land\; \left( r_1<1-\frac{m_{c}}{M}
  \right), \\
  &A_1:(1-r_1)(1-r_2)<\frac{m_{c}}{M}, A_2:(1-r_1)r_2<\frac{m_{c}}{M}, \\
  &A_3:(1-r_2)r_1<\frac{m_{c}}{M}, A_4:r_1r_2<\frac{m_{c}}{M}.
\end{eqnarray}
With these events the probability can be expressed as,
\begin{eqnarray}
  P^{fb}_2&=&\frac{1}{2} \left\{P(A_1 \lor A_2 |A_0) + P(A_3 \lor A_4
    |A_0) \right\} \nonumber \\ 
  &=&\frac{1}{2} \left\{ 
    P(A_1|A_0)+P(A_2|A_0)-P(A_1 \land A_2|A_0)\right.  \nonumber \\ 
   &&\hspace{-0.3cm}\left.+P(A_3|A_0)+P(A_4|A_0)-P(A_3 \land A_4|A_0) \right\},
\end{eqnarray}
where the factor $1/2$ reflects the two existing subcases for the two
fireballs formation case as it is shown in \figref{Fig:2fireball}. We
carried out the calculations -- based on geometric probabilities --
for the one, two and three fireballs formation, however, the
probability has a closed form only in the first two cases, which are
\begin{eqnarray}
 \hspace*{-0.5cm} P^{fb}_1(M)&=&
  \begin{cases}
    \frac{2m_{c}}{M},\; \hfill  0<\frac{m_{c}}{M}<\frac{1}{2} \\
    1,\; \hfill \frac{1}{2}<\frac{m_{c}}{M}<1
  \end{cases},\label{Eq:Pfb1}\\
\hspace*{-0.5cm}  P^{fb}_2(M)&=&
  \begin{cases}
    \frac{2m_{c}}{M}\left[ \ln\left( {\frac{M}{2m_{c}}-\frac{1}{2}} \right) + \frac{1}{2}\right], 
    \;\hfill 0<\frac{m_{c}}{M}<\frac{1}{3}\\ 
    1-\frac{2m_{c}}{M},\; \hfill \frac{1}{3}<\frac{m_{c}}{M}<\frac{1}{2}
  \end{cases}\hspace*{-0.4cm},\label{Eq:Pfb2}
\end{eqnarray}
while in the third case it can be found in \apref{App:fr3} due to
its lengthy formula. Important consequence are the ratios of the
fireball formation probabilities $P^{\text{fb}}_{k}$, since they
clearly show that if the invariant mass -- or collision energy -- is
increased, it is more likely to get more and more fireballs. Thus more
particles can be produced. The ratio of
$P^{\text{fb}}_{2}/P^{\text{fb}}_{1}$ can be seen in
\figref{Fig:P2overP1}, where the result of a Monte\,-\,Carlo
simulation is also shown.

The last ingredient in \eref{Eq:W_calc} is the the
quark\,-\,com\-bi\-na\-to\-ri\-al factor $C_Q(M)$, which is a
dynamical -- that is energy dependent -- parameter of the model and is
necessary to describe baryonic final states. A similar approach can be
found in \cite{Cuautle:2013}, where the abundances of baryons
containing strange quarks was described by quark combinatorial
probabilities. The basic assumption is that the probability of the
creation of a specific hadron is proportional to the number of ways
that the hadron can be formed from a given number of $u,d,b,s,c,t$
quark\,-\,antiquark pairs. Thus, it is necessary to know as a function
of energy/invariant mass the number of quark\,-\,antiquark pairs and
the distribution of the various quark species once the average number
of pairs is known. The number of quark\,-\,antiquark pairs is
calculated in \cite{Degtyarenko:2000} based on phase space
considerations, which reads
\begin{equation}
  N(M)=\frac{1+\sqrt{1+M^2/T_0^2}}{2},
\end{equation}
where $T_0$ is the interaction temperature, which is found to be
$T_0=160$~MeV, the Hagedorn temperature. It should be noted that only
$u,d,s,c$ quark\,-\,antiquark pairs are considered in our
investigation. It is assumed that the light up and down
quark\,-\,antiquark pairs are formed with the same $P_u=P_d$
probability due to their almost equal mass, but heavier pairs such as
the strange and charm have suppression in their creation probabilities
$P_s,P_c < P_u$. These probabilities are important if there are
strange or charm particles in the final state such as $K$ or $J/\Psi$.
The $P_s$ and $P_c$ quark\,-\,antiquark pair probabilities are fitted
to experimental data, which is discussed in
\secref{Subsec:final_sc}. It is worth to note that in the four quark
approximation the pair formation probabilities should satisfy the
$P_u+P_d+P_s+P_c=1$ relation. Once the pair probabilities are known
we also need a probability mass function $F(n_u,n_d,n_s,n_c;N(M))$,
which gives the probability that from the $N(M)$ quark\,-\,antiquark
pairs we have exactly $n_u, n_d, n_s, n_c$ number of $u\bar u$, $d\bar
d$, $s\bar s$, $c\bar c$ pairs, respectively. We used the multinomial
distribution that has the following probability mass function,
\begin{equation}
  F(n_u,n_d,n_s,n_c;N(M)) = \frac{\left( \frac{N(M)!}{n_u! n_d! n_s!
      n_c!} \right) P_u^{n_u} P_d^{n_d}P_s^{n_s}P_c^{n_c}}{P_{tot}(N)},
  \label{Eq:quark_distrF}
\end{equation}
where $P_{tot}(N)$ is the normalization factor, which sums the
nominator over all the possible $\{n_u,n_d,n_s,n_c\}$ combinations
that satisfy the $n_u+n_d+n_s+n_c=N$ constraint. It is very important
to note that we assumed that such $n_u,n_d, n_s, n_c$ values will be
realized that maximize $F(n_u,n_d,n_s,n_c;N(M))$.

\begin{figure}[!t]
    \includegraphics[width=0.5\textwidth]{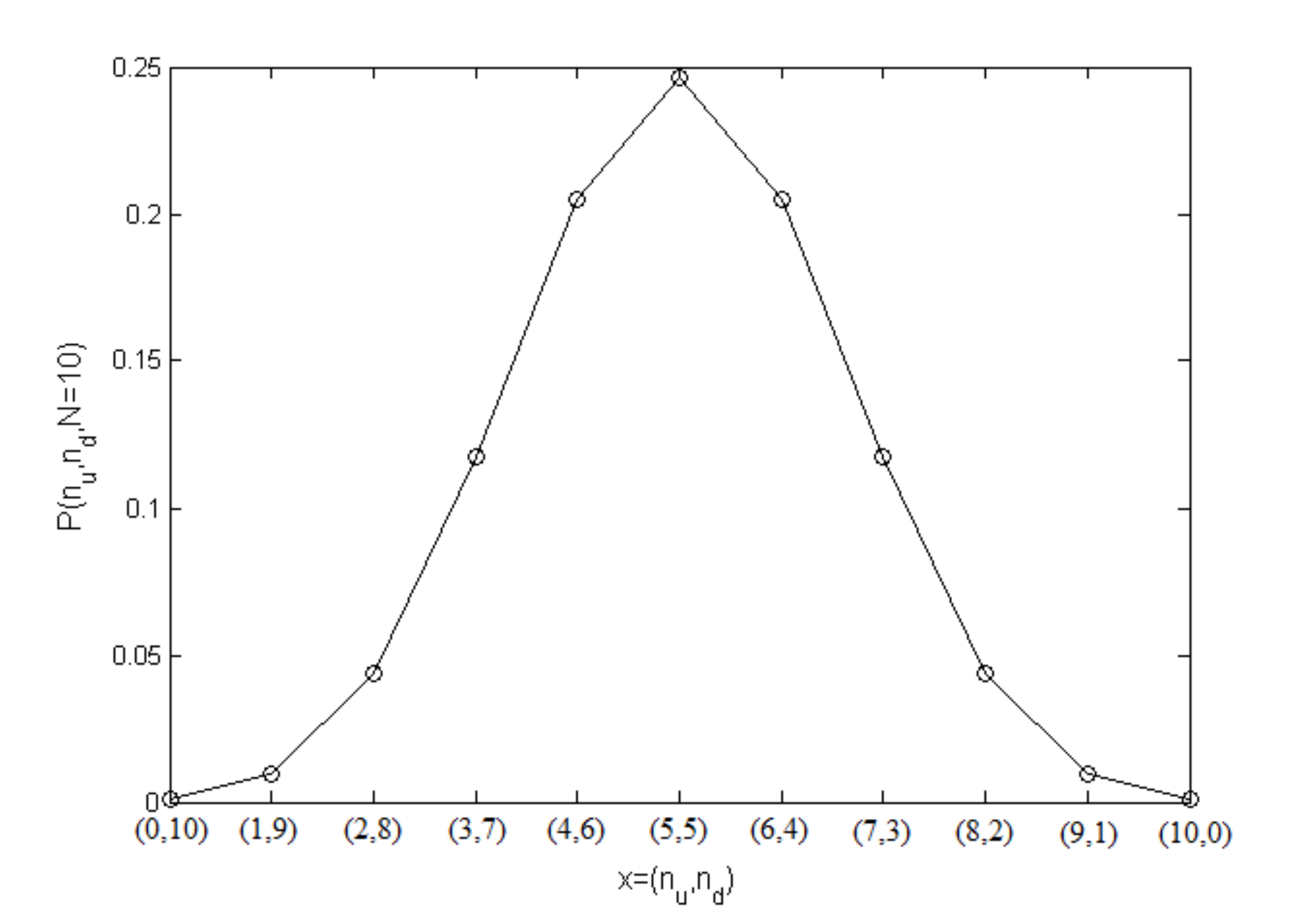}
    \caption{The distribution function in the two quark case for
      $N=10$ from Monte\,-\,Carlo simulation. On the $x$\,-\,axis
      $\{n_u,n_d\}=\{n_u,N-n_u\}$ is shown.}
    \label{Fig:prob_ud_N_10}  
\end{figure}

Suppose for a moment that we have only $u$ and $d$ quarks, then since
$n_u+n_d=N$ and $P_u+P_d=1$, $F(n_u,n_d=N-n_u;N) = F(n_u;N)$ and
$P_u=P_d=0.5$. One particular case, $N=10$, can be seen in
\figref{Fig:prob_ud_N_10}.
It can be seen that the distribution is symmetric and have a unique
maximal value, which is the case with more quarks as
well. Consequently, by using the values given by the maximum of
\eqref{Eq:quark_distrF} the energy dependent quark\,-\,combinatorial
factors can be easily written for the various final states. For
instance, in case of $M($fireball$)\rightarrow p(uud)+n(udd)$ one has,
\begin{equation}
  \label{Eq:CQ}
  \begin{split}
    C_{Q, pn}(M)=P_{(u,d,s,c)}^{\text{max}}& 3^2
    n_u(n_u-1)n_d\\
    &\times(n_u-2)(n_d-1)(n_d-2),
  \end{split}
\end{equation}
where $P_{(u,d,s,c)}^{\text{max}} \equiv
\max\left(F(n_u,n_d,n_s,n_c;N(M)) \right)$ is the maximal value of the
probability that $n_u,n_d,n_s,n_c$ number of quark\,-\,antiquark pairs
were created, and the $3^2$ factor is for the number of colorless
combinations of the two baryons. Connected to \eref{Eq:CQ} it should
be noted that currently we do not care about the fate of the created
antiquarks or quarks, which are not used for any hadrons. That is for
instance in \eref{Eq:CQ} we disregard three $\bar u$'s and three $\bar
d$'s. We plan to address this issue in a subsequent work.

To get $C_Q(M)$ the final step is to calculate a normalization factor
according to the multiplicity of the decay (two or three body):
\begin{equation}
   \label{Eq:CQTOT}
   C_Q(M) = C_{Q,pn}\mathcal{N_{C}}_{,2}(M) 
\end{equation} 
In the normalization factor we have to sum over all the possible quark
combinations that is allowed by the conservation laws of the fireball.
In the cases of two and three body decays the normalizations are
\begin{equation}
   \label{Eq:CQNORM1}
   \mathcal{N_{C}}_{,2}(M) = \left[ \sum_{<ij> \in S} C_{Q,(ij)} \right]^{-1},
\end{equation} 
\begin{equation}
   \label{Eq:CQNORM2}
   \mathcal{N_{C}}_{,3}(M) = \left[ \sum_{<ijk> \in S} C_{Q,(ijk)} \right]^{-1},
\end{equation} 
where $S$ represents the set of all possible hadron combinations that
has the same quantum numbers as the fireball. With this normalization
$C_Q(M)$ can be considered as a quark combinatorial probability. In
the two fireball case we will have two combinatorial factors, hence
the final combinatorial probability will be $C_Q = C_{Q_1}C_{Q_2}$.
For other final states $C_Q$ can be calculated similarly. In our model
$P_{(u,d,s,c)}^{max}$ and the number of quark\,-\,antiquark pairs
$n_i$ are also energy dependent. This will be important when the pair
creation probability is very small, e.g. channels with charm quarks.

Finally, let us give the explicit form of the total formation
probability $W$ in the cases of one and two fireball formation. As was
already discussed, one fireball can decay into two or three hadrons,
thus $n_i=2,3$ in $P^{H,i}_{n_i}$, while the number of outgoing
hadrons can be two or three in case of one fireball, while four to six
in case of two fireballs. The total probabilities are
\begin{eqnarray}
  W_1^{2,3}(M) &=& \mathcal{N}_1(M)P^{fb}_1(M) C_Q(M) P^{H}_{2,3}(M),\label{Eq:W1}\\
  W_2^{2,3;2,3}(M) &=& \mathcal{N}_2(M) P^{fb}_2(M)
  C_Q(M)\label{Eq:W2}\nonumber\\
  &&\times\int_{x_{min}}^{x_{max}}dx P^{H,1}_{2,3}(M-x) P^{H,2}_{2,3}(x),
\end{eqnarray}
where $C_Q(M)$, as was already discussed, depends on the considered
final state, and $P^{fb}_{1,2}$ and $P^{H}_{2,3}$ are given by
Eqs. \eqref{Eq:Pfb1}, \eqref{Eq:Pfb2} and Eqs.~\eqref{Eq:PH2},
\eqref{Eq:PH3}, respectively. Moreover, the integration limits
$x_{min}$ and $x_{max}$ have to respect the kinematic limits of the
final states. For instance in case of $n_1=n_2=2$ -- four hadron final
state -- with $\text{fireball}_1 \to a + b$ and $\text{fireball}_2 \to
c + d$ decay scheme one has $x_{min}=m_c+m_d$ and $x_{max}=M-m_a-m_b$.

For a few simple processes with low multiplicity -- e.g. two or three
particles -- the probabilities, and thus the ratios of probabilities
of different processes can be expressed analytically, however, for
more complicated processes numerical methods are necessary. In the
following section we show that the model is able to describe low
energy cross sections and their ratios with good accuracy.

\section{Results}
\label{Sec:result}

\subsection{Final states with up and down quarks}
\label{Subsec:final_ud}

\begin{figure}[!t]
    \includegraphics[width=0.45\textwidth]{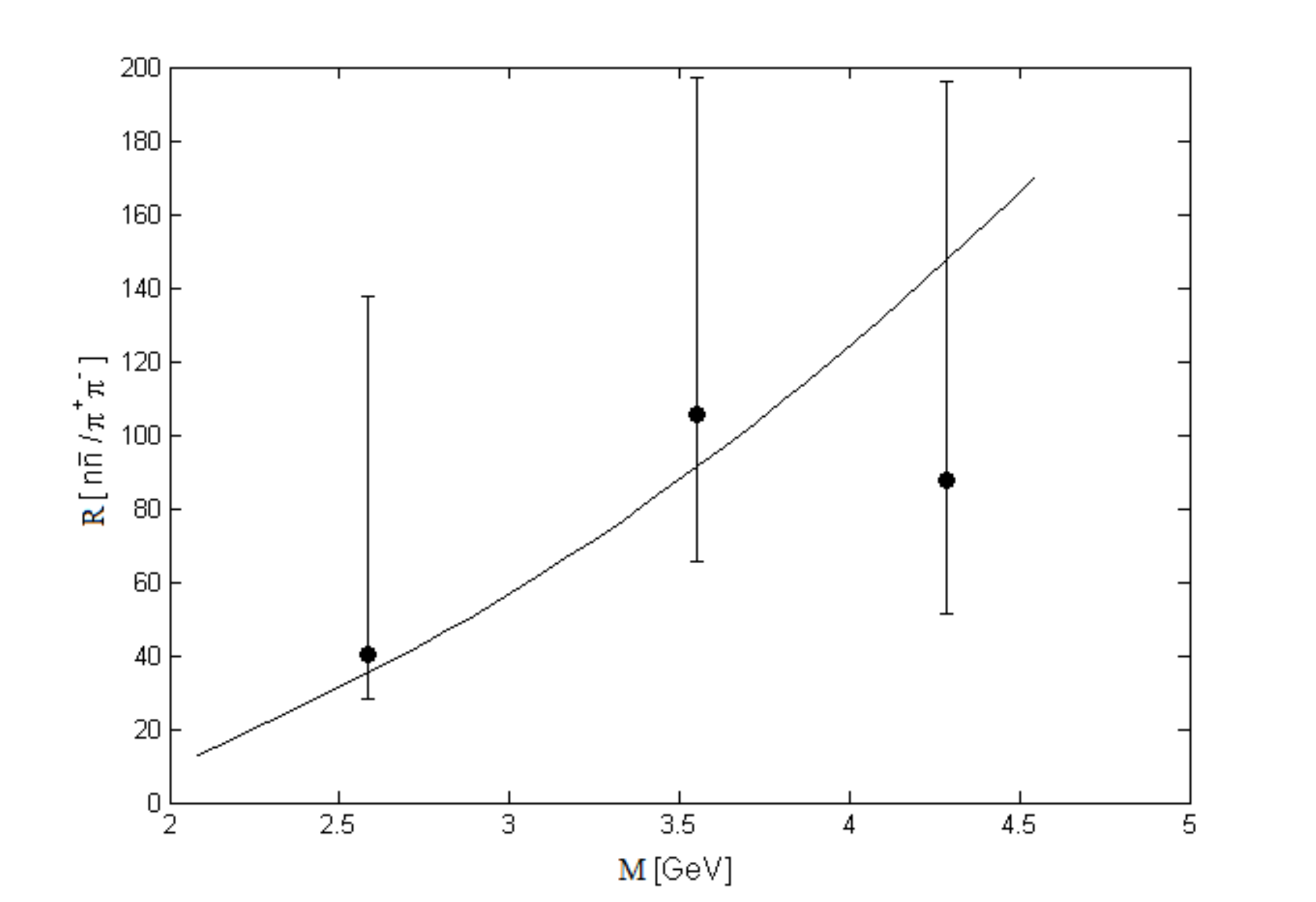}
    \caption{The ratio $R^{n\bar n}_{\pi^+\pi^-}$ -- see
      \eref{Eq:R_nn_pipi} -- as a function of $M$ (solid line)
      together with the measured data (filled circles) from
      \cite{Baldini:1988}}
    \label{Fig:nn_per_pipi}  
\end{figure}

The simplest processes that can be calculated are the ones with two
particles in the final state. In this subsection only the lightest u
and d quarks are considered and for their formation probabilities
$P_u=P_d=0.5$ is assumed. As a first example we would like to
reconstruct the following ratio
\begin{equation}
  R^{n\bar n}_{\pi^+\pi^-}(M) = \left.\left( \frac{\sigma_{p\bar p\rightarrow n\bar
      n}}{\sigma_{p\bar p}^{Tot}} \right) \middle/
  \left( \frac{\sigma_{p\bar p\rightarrow \pi^+ \pi^- }}{\sigma_{p\bar p}^{Tot}}\right)\right. .
\end{equation}
Consequently, $W_{n\bar n}$ and $W_{\pi^+\pi^- }$ should be calculated
from the model. Since the two\,-\,particle final states can only come
from a one\,-\,fireball decay, \eref{Eq:W1} should be used. The quark
combinatorics can be given based on \eref{Eq:CQ} and on the quark
content of the final state, which is $n \sim udd$, $\bar n \sim \bar u\bar d \bar d$,
$\pi^{+} \sim u\bar d$, $\pi^{-} \sim \bar u d$; and
the combinatorial factors are,
\begin{eqnarray}
  C_{Q,n\bar n}&=&3^2 P_{u,d}^{max} n_d^2 (n_d-1)^2 n_u^2,\\
  C_{Q,\pi^+ \pi^- }&=&3^2 P_{u,d}^{max} n_u^2 n_d^2, \label{Eq:CQpi+pi-}  
\end{eqnarray}
where $n_u=n_d=N/2$, since these values give the maximum of the
probability mass function $F(n_u,n_d;N(M))$, and $P_{u,d}^{max} =
F(n_u=N/2,n_d=N/2;N)$. The resulting expression will be quite simple
due to the common factors that cancel out and it reads
  \begin{equation}
    \label{Eq:R_nn_pipi}
    \begin{split}
      R^{n\bar n}_{\pi^+\pi^-}(M) &= \frac{W_{n\bar n}(M)}{W_{\pi^+\pi^-}(M)}=4 \left(
      \frac{1+\sqrt{1+M^2/T_0^2 }}{4} - 1\right)^2 \\ & \times \left(
      \frac{M^2-4m_{n}^2}{M^2-4m_{\pi}} \right)^{1/2},
    \end{split}
  \end{equation}
 where $m_n$ and $m_{\pi}$ are the masses of the neutron and pion,
 respectively. The result together with the measured data can be seen
 in \figref{Fig:nn_per_pipi}, which shows a good agreement in the
 given energy range, however, the error bars are quite large.

\begin{figure*}[!t]
    \centerline{\includegraphics[width=0.75\textwidth]{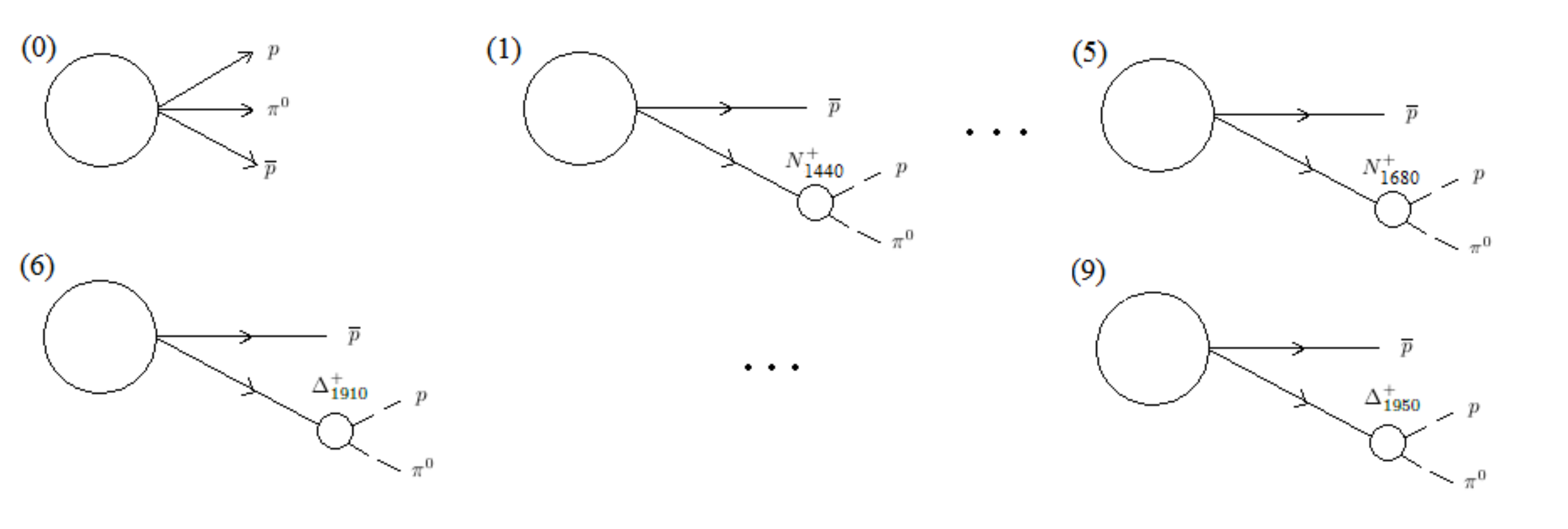}}
    \caption{Possible one\,-\,fireball decay schemes for the $p\bar
      p\rightarrow p\bar p \pi^0$ process. Similar set of diagrams can
      be shown for the antiresonances, where the protons and
      antiprotons should also be exchanged.}
    \label{Fig:dec_sch_pap_pappi}  
\end{figure*}

\begin{figure}[!t]
    \includegraphics[width=0.45\textwidth]{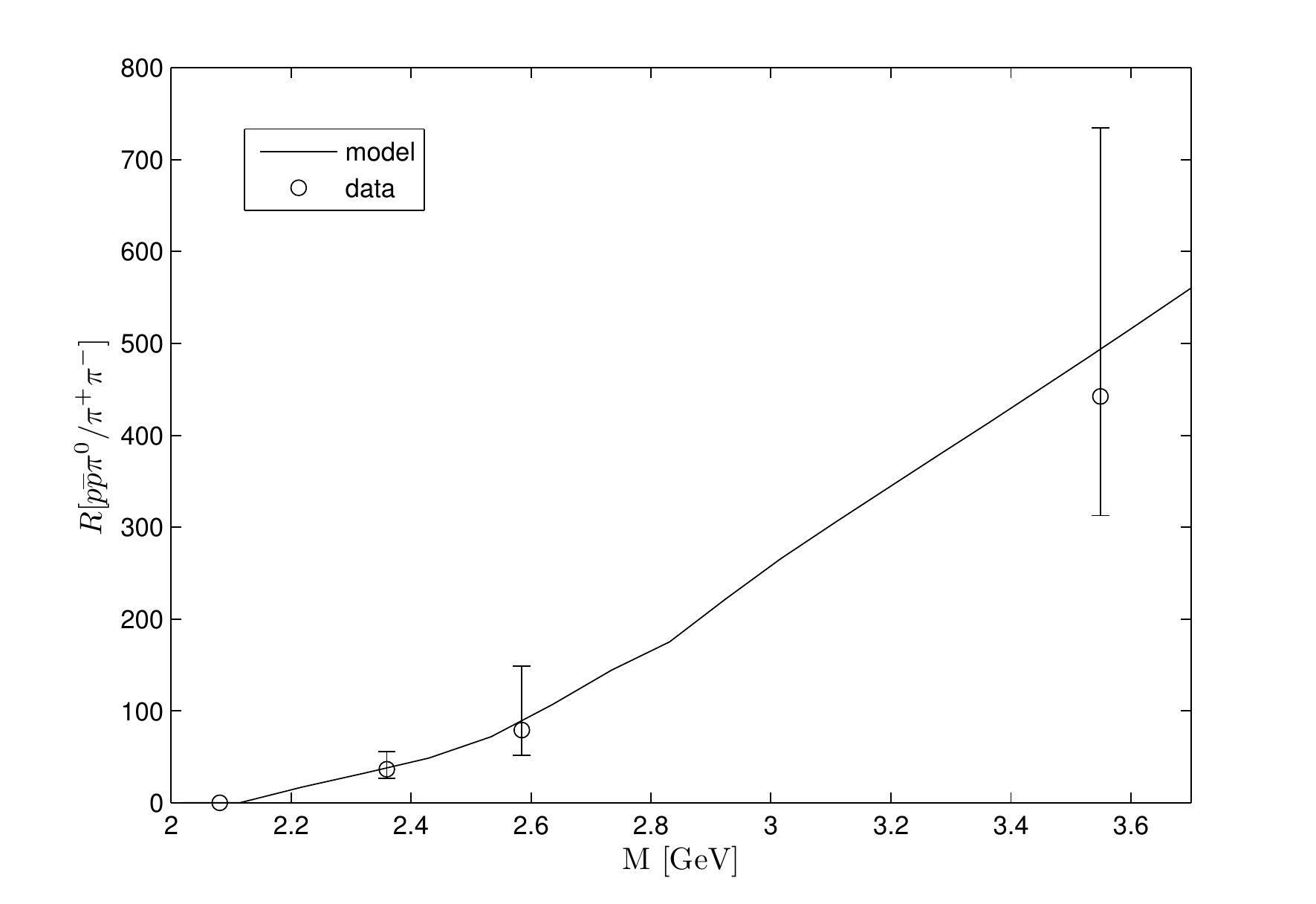}
    \caption{The ratio $R^{p\bar p\pi^0}_{\pi^+\pi^-}$ -- see
      \eref{Eq:R_pppi0_pipi} -- as a function of $M$ (solid line)
      together with the measured data (circles) from
      \cite{Baldini:1988}}
    \label{Fig:pppi0_per_pipi}  
\end{figure}

As another example we calculated the ratio of the cross sections of
the reactions $(p\bar p\rightarrow p\bar p \pi^0)$ and $(p\bar p
\rightarrow \pi^+ \pi^-)$, which is slightly more complicated than the
previous one, because the $p\pi^0$ two\,-\,particle final state can
arise from many resonances. Thus, we have to take into account all the
possible $p\bar p \to R\bar p$, or $p\bar p \to \bar R p$ final states
as well, where $R$ can be any resonance which decays into
$p\pi^0$. The possible decay schemes can be seen in
\figref{Fig:dec_sch_pap_pappi}.
It is worth to note that in the figure the zeroth scheme corresponds
to a three particle hadronization process, while all the rest to two
particle hadronization processes.  In this calculation, such nucleon
and delta resonances are considered -- taken from PDG \cite{PDG} --
that has a larger branching ratio than $15\%$ for the $p\pi^0$
channel. These are listed in \tabref{Tab:N_D_resonances} together with
their masses, spins and branching ratios of the $p\pi^0$ channel
$B_i^{p\pi^0}$. We considered resonances and antiresonances alike.
\begin{table}
\caption{Considered nuclear and delta resonances their masses
  $m_{R_i}$, spins $s_{R_i}$ and branching ratios $B_i^{p\pi^0}$. We
  excluded resonances with $B_i^{p\pi^0} < 0.15$. The same values
  apply to the corresponding antiresonances.}
\label{Tab:N_D_resonances}
\begin{center}
\begin{tabular}{llccc}
\hline\noalign{\smallskip}
i & $R_i$ & $m_{R_i}$ [GeV] & $s_{R_i}$ & $B_i^{p\pi^0}$\\
\noalign{\smallskip}\hline\noalign{\smallskip}
$1$ & $N_{1440}$      & $1.430$ & $1/2$ & $0.22$ \\
$2$ & $N_{1520}$      & $1.515$ & $3/2$ & $0.20$ \\
$3$ & $N_{1535}$      & $1.535$ & $1/2$ & $0.15$ \\                         
$4$ & $N_{1650}$      & $1.655$ & $1/2$ & $0.23$ \\
$5$ & $N_{1680}$      & $1.685$ & $5/2$ & $0.23$ \\
$6$ & $\Delta_{1232}$ & $1.232$ & $3/2$ & $0.66$ \\ 
$7$ & $\Delta_{1620}$ & $1.630$ & $1/2$ & $0.17$ \\ 
$8$ & $\Delta_{1910}$ & $1.890$ & $1/2$ & $0.15$ \\ 
$9$ & $\Delta_{1950}$ & $1.930$ & $7/2$ & $0.27$ \\
\noalign{\smallskip}\hline
\end{tabular}
\end{center}
\end{table}
The necessary quark combinatorial factors for this ratio are
  \begin{eqnarray}
    C_{Q,p\bar p \pi^0}&=&3^3 P_{u,d}^{max} n_u^2 (n_u-1)^2 n_d^2\nonumber\\
    &\times&\frac{(n_u-2)^2 + (n_d-1)^2}{2},\\
  C_{Q,\bar p R}&=& C_{Q,p\bar R} = 3^2 P_{u,d}^{max} n_u^2 (n_u-1)^2 n_d^2,      
  \end{eqnarray}
while $C_{Q,\pi^+ \pi^- }$ is already given in \eref{Eq:CQpi+pi-}.
After some manipulation we find,
\begin{eqnarray}
  R^{p\bar p\pi^0}_{\pi^+\pi^-}(M)&=&\frac{W_{p\bar p\pi^0}}{W_{\pi^+\pi^-}}=\frac{(n_u-1)^2}{\Phi_2(M ,m_{\pi},m_{\pi})}\nonumber\\
  &\times&\Bigg[\frac{6P^d_3}{P^d_2(2\pi)^3}\frac{\mathcal{N_C}_{,3}}{\mathcal{N_C}_{,2}}\left[(n_u-2)^2 + (n_d-1)^2\right]\nonumber\\
    &\times&\Phi_3(M ,m_p,m_p,m_{\pi^0}) + 4\sum_{i=1}^9B_i^{p\pi^0}(2s_{R_i}+1)\nonumber\\
    &\times&\Phi_2(M,m_p,m_{R_i}) \Bigg],\label{Eq:R_pppi0_pipi}
\end{eqnarray}
where $m_{p}$ and $m_{R_i}$ are the masses of the proton and the
resonances, respectively, while the explicit form of the phase space
integrals $\Phi_{2,3}$ can be found in \apref{App:phi23}. Moreover,
the factor of $4$ in front of the last term comes form two source: a
factor of $2$ from the summation over the antiresonances, and another
factor of $2$ from the proton spin degeneration. The ratio as a
function of energy/invariant mass and its comparison with experimental
data can be seen in \figref{Fig:pppi0_per_pipi}, where a remarkably
good match was found.

\subsection{Final states with strange and charm quarks}
\label{Subsec:final_sc}

\begin{figure}[!t]
    \includegraphics[width=0.45\textwidth]{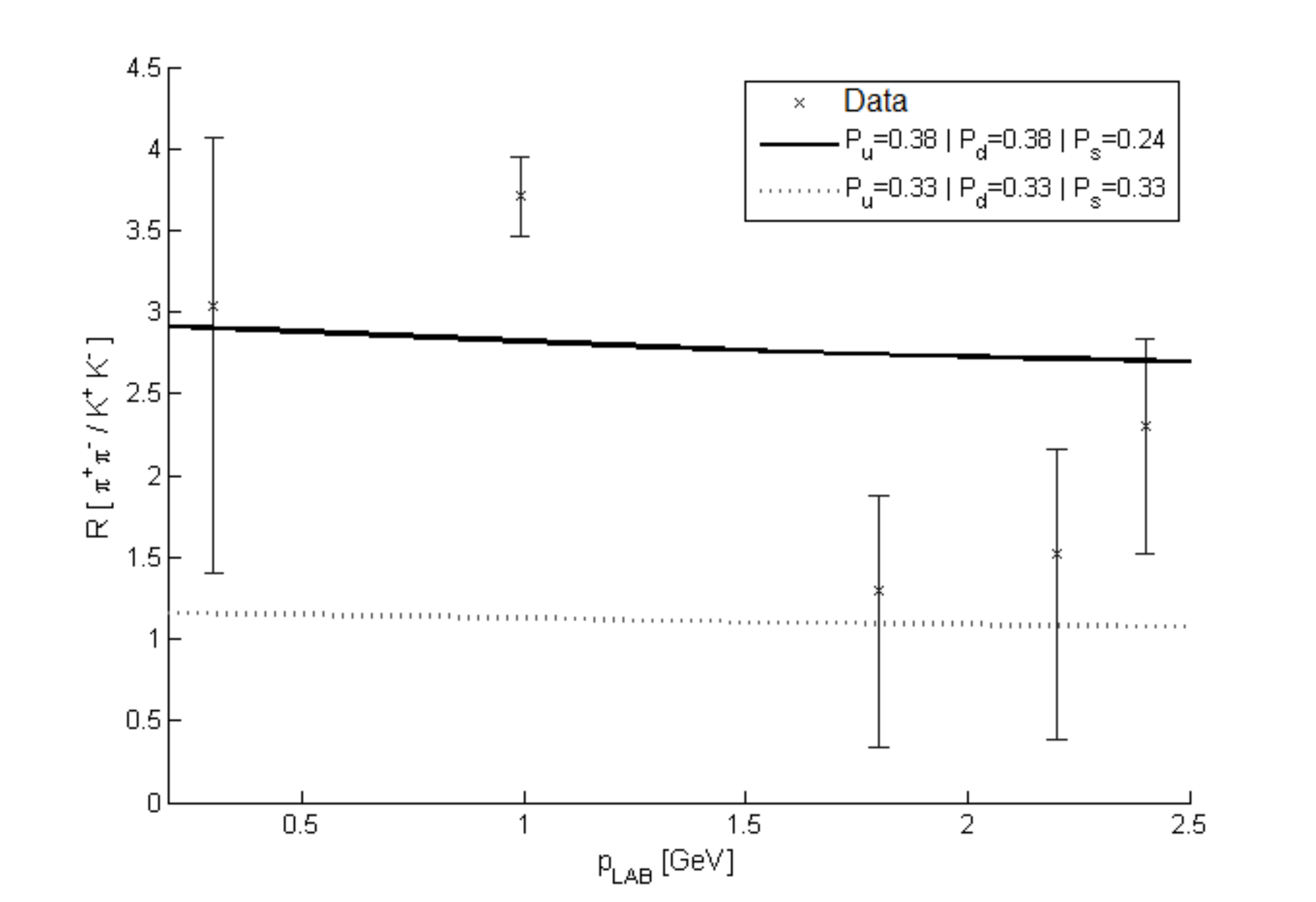}
    \caption{Calculated values and experimental data of the ratio
      $R^{\pi^+ \pi^-}_{K^+K^-}(M)$. Dotted line correspond to the
      equal probability case $P_u=P_d=P_s$, while the solid line to
      the strange suppression scenario.}
    \label{Fig:R_pipi_KK}  
\end{figure}

For final states that also contain heavier quarks -- in our case
strange or charm -- the assumption of equal quark creation
probabilities is not a good approximation \cite{Park:2014}. In this
case one can fit the creation probability parameters $P_s$ and $P_c$
to existing data. Besides, it is also possible to estimate these
parameters from some theory (see e.g. the Hawking\,-\,Unruh
hadronization model \cite{Castorina:2014_2015}). In order to determine
the strange quark probability $P_s$, we calculated the ratio
$\frac{\sigma_{p \bar p\to \pi^+ \pi^-}}{\sigma_{p\bar p \to K^+K^-}}$
in the energy range $E \in \{1.876,2.602\}$~GeV, which has a similar
expression as Eq.~\eqref{Eq:R_nn_pipi} and it reads
\begin{equation}
  R^{\pi^+ \pi^-}_{K^+K^-}(M)=\frac{n_d^2}{n_s^2} \left(\frac{M^2-4m_{\pi}^2}{M^2-4 m_{K}^2 } \right)^{1/2},
  \label{Eq:R_pipi_KK}
  \end{equation}
 where $m_K$ is the kaon mass. As it can be seen,
 $R^{\pi^+\pi^-}_{K^+K^-}(M)$ strongly depends on the $n_d/n_s$ ratio,
 which was adjusted through the change of $P_u$ and $P_s$ to get the
 best agreement with the measured data. We used the following
 iterative procedure, first we set some initial values for $P_u$ and
 $P_s$ then calculated the maximum of the probability mass function
 $F$ in \eref{Eq:quark_distrF}, which gave the values of $n_u$ and
 $n_s$. With these values we calculated the ratio in
 \eref{Eq:R_pipi_KK} and compared with the experimental data, then
 changed the $P_u$ and $P_s$ and recalculated the ratio $R^{\pi^+
   \pi^-}_{K^+K^-}(M)$. We repeated this process until we got the
 smallest deviation form the experimental data. It turns out that we
 get the well\,-\,known strange quark suppression
 \cite{Becattini:1997a,Drescher:2001,Sjostrand:1993}. In
 \figref{Fig:R_pipi_KK} the original equal probability case
 $P_u=P_d=P_s=1/3$ can be seen together with the fitted strange
 suppression case, where we get for the probabilities, $P_u=P_d=0.38$,
 and $P_s=0.24$. In the figure the experimental data from
 \cite{Baldini:1988} marked with crosses and error bars is also shown.
It can be seen that when the equal probability assumption is used
(dotted line) the calculated ratio underestimates by an approximate
factor of three the measured one, which means that the strange channel
is overestimated by the model. When the strange quark creation
probability was decreased, however, a better agreement could be
achieved (solid line). In conclusion we can say that inclusion of an
extra suppression parameter is not necessary in our approach, but in
exchange we have to tune the $P_u,P_d,P_s$ probability parameters.
The previous example of fitting $P_s$ was merely pedagogical due to
the large errors and the few data points. Other recent fits -- not
discussed here -- using channels involving $\Lambda$ and $\bar\Lambda$
particles shows us that $P_s$ could have a little smaller, while $P_u,
P_d$ a little higher values, however, it is necessary to include more
channels to make a reliable fit. Another important point to note is
that for a wider energy range $P_s$ should be energy dependent. This
is due to the natural assumption that at very large energy
$P_u=P_d=P_s$ should hold. For the energy depend-ant suppression
factor please see \cite{Satz:2016aty}.

This technique can be applied to determine the charm quark creation
probability $P_c$. Subsequently, $P_c$ can be used to estimate cross
sections with final states containing charm quarks, like for instance
to the reaction $p+\bar p\to \pi^0+J/\Psi$. It should be noted,
however, that the existing measured data are still very limited, thus
only a crude estimation can be made.

\begin{figure}[!t]
    \includegraphics[width=0.45\textwidth]{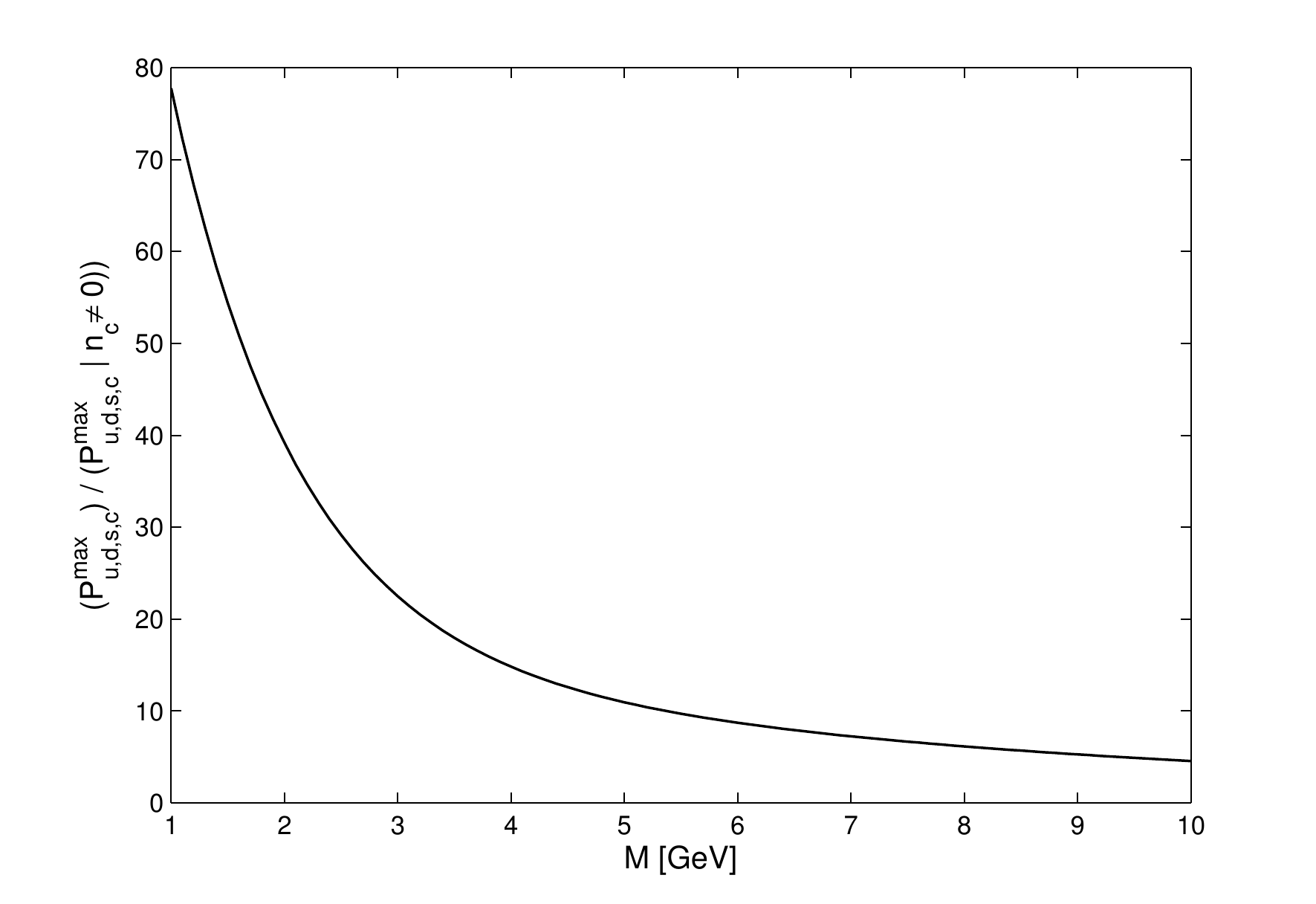}
    \caption{$M$ dependence of the ratio $r_P$ of the probabilities
      appearing in \eref{Eq:R_pp_J} (see text for details).}
    \label{Fig:R_pp_J}  
\end{figure}

To determine $P_c$ we considered the inclusive reaction $(p + p \to p
+ p + J/\Psi + X)$ in the low energy regime and used a parametrization
of the $\sigma_{J/\Psi}^{pp}$ cross section -- based on experimental
data -- taken from \cite{Linnyk:2006}. Since calculation of the
inclusive reaction probability is rather involved, we used the data
near the $(p+p+J/\Psi)$ threshold instead, where no other particles
are created. For the fit we calculated the ratio
$R_{ppJ/\Psi}^{pp\pi^0}=W_{pp\pi^0}/W_{ppJ/\Psi}$ and adjusted the
result -- by changing the quark pair creation probabilities $P_u$,
$P_d$, $P_s$, and $P_c$-- to the measured ratio
$\sigma_{pp\pi^0}/\sigma_{ppJ/\Psi}$ near the threshold, {\it i.e.}
$\sqrt{s}\equiv M =5$~GeV. Since the charm quark mass $1.28$~GeV is
much larger than the $u, d, s$ quark masses $2$~MeV, $5$~MeV, and
$95$~MeV, respectively, it is expected that its pair creation
probability $P_c$ will be much lower. If we simply used in $C_Q$, as
we did previously, the maximum value $P_{u,d,s,c}^{\text{max}}$ of the
probability mass function $F$, then most probably we would get zero
$c$ quarks, which was checked explicitly. Obviously that maximum for
$P_{u,d,s,c}$ can not be used to calculate the $W_{ppJ/\Psi}$
probability. Thus for a reaction, where the final state involves charm
quark(s) we use the conditional maximum
$P_{u,d,s,c}^{\text{max}}|_{n_c\ne 0}$ instead, while for other
reactions we retain the original $P_{u,d,s,c}^{max}$. The calculated
ratio reads as,
  \begin{eqnarray}
  \label{Eq:R_pp_J} 
    &&R_{ppJ/\Psi}^{pp\pi^0}(M)=\frac{P_{u,d,s,c}^{\text{max}}}{P_{u,d,s,c}^{\text{max}}|_{n_c\ne
        0}} \frac{1}{n_c^2}\Bigg[ \left((n_u-4)n_u\right.
      \nonumber \\
    &&\left.+(n_d-2)n_d\right) \frac{\Phi_3(M,m_p,m_p,m_{\pi})}{2} + 
      \frac{P^d_2(2\pi)^3}{3P^d_3} \frac{\mathcal{N_C}_{,2}}{\mathcal{N_C}_{,3}} \nonumber \\
    &&\times\sum_{i=1}^9 \Big( B_i^{p\pi^0} (2s_{R_i}+1) \Phi_2(M,m_p,m_{R_i}) \Big)\Bigg] \Bigg/
      \sum_{i=1}^3 \Big(B_i^{J}   \nonumber \\
    && \times (2s_{R^J_i}+1) \Phi_3(M,m_p,m_p,m_{R^J_i}) \Big),
  \end{eqnarray}
 where $m_{R_i}$, $B_i^{p\pi^0}$, and $s_{R_i}$ are the masses,
 branching ratios and spins of the nucleon and delta resonances
 already given in \tabref{Tab:N_D_resonances}, while $m_{R^J_i}$,
 $B_i^{J}$ and $s_{R^J_i}$ are the masses, branching ratios -- of
 $J/\Psi$ resonances decaying into $J/\Psi$ -- and spins of the
 $J/\Psi$ resonances and are listed in \tabref{Tab:J_resonances}. The
 branching ratios were taken from \cite{Barnes:2006}.
\begin{table}
\caption{Considered $J/\Psi$ resonances their masses $m_{R^J_i}$,
  spins $s_{R^J_i}$ and branching ratios $B_i^{J}$}
\label{Tab:J_resonances}
\begin{center}
\begin{tabular}{llccc}
\hline\noalign{\smallskip}
i & $R_i$ & $m_{R^J_i}$ [GeV] & $s_{R^J_i}$ & $B_i^{J}$\\
\noalign{\smallskip}\hline\noalign{\smallskip}
$1$ & $\Xi_{c1}$      & $3.511$ & $1$ & $0.30$ \\
$2$ & $\Xi_{c2}$      & $3.556$ & $2$ & $0.30$ \\
$3$ & $\Psi^{\prime}$  & $3.686$ & $1$ & $0.56$ \\                         
\noalign{\smallskip}\hline
\end{tabular}
\end{center}
\end{table}
Based on \cite{Linnyk:2006,Barnes:2006} the experimental value for the
ratio at threshold is $R_{ppJ/\Psi}^{pp\pi^0}(M=5\text{GeV}) \approx 8.73 \cdot 10^{7}$.
After fitting our calculated ratio -- similarly as in the case of
$P_s$ -- to the experimental value, it was found for the
quark\,-\,antiquark creation probabilities
\begin{equation}
  \label{Eq:Pudsc_values}
  P_u = P_d = 0.38, P_s=0.2386, P_c=0.0014
\end{equation}
It turns out that in this energy range the $n_c=1$ condition gives the
local maximum $P_{u,d,s,c}^{\text{max}}|_{n_c\ne 0}$, thus in the quark
combinatorial factor the $n_c=1$ value was used. With these values we
calculated the invariant mass/energy dependence of the probability
ratio $r_P \equiv P_{u,d,s,c}^{\text{max}} /
P_{u,d,s,c}^{\text{max}}|_{n_c\ne 0}$ appearing in \eref{Eq:R_pp_J},
which can be seen in \figref{Fig:R_pp_J}.
The ratio of the global and the local maximums tends to unity (not
shown explicitly), which means that at higher energies -- where the
ratio is close to $1$ -- only the global maximums of the probability
mass distribution function is needed as the number of $c\bar c$ pairs
will always be larger than zero.

\begin{figure}[!t]
    \includegraphics[width=0.45\textwidth]{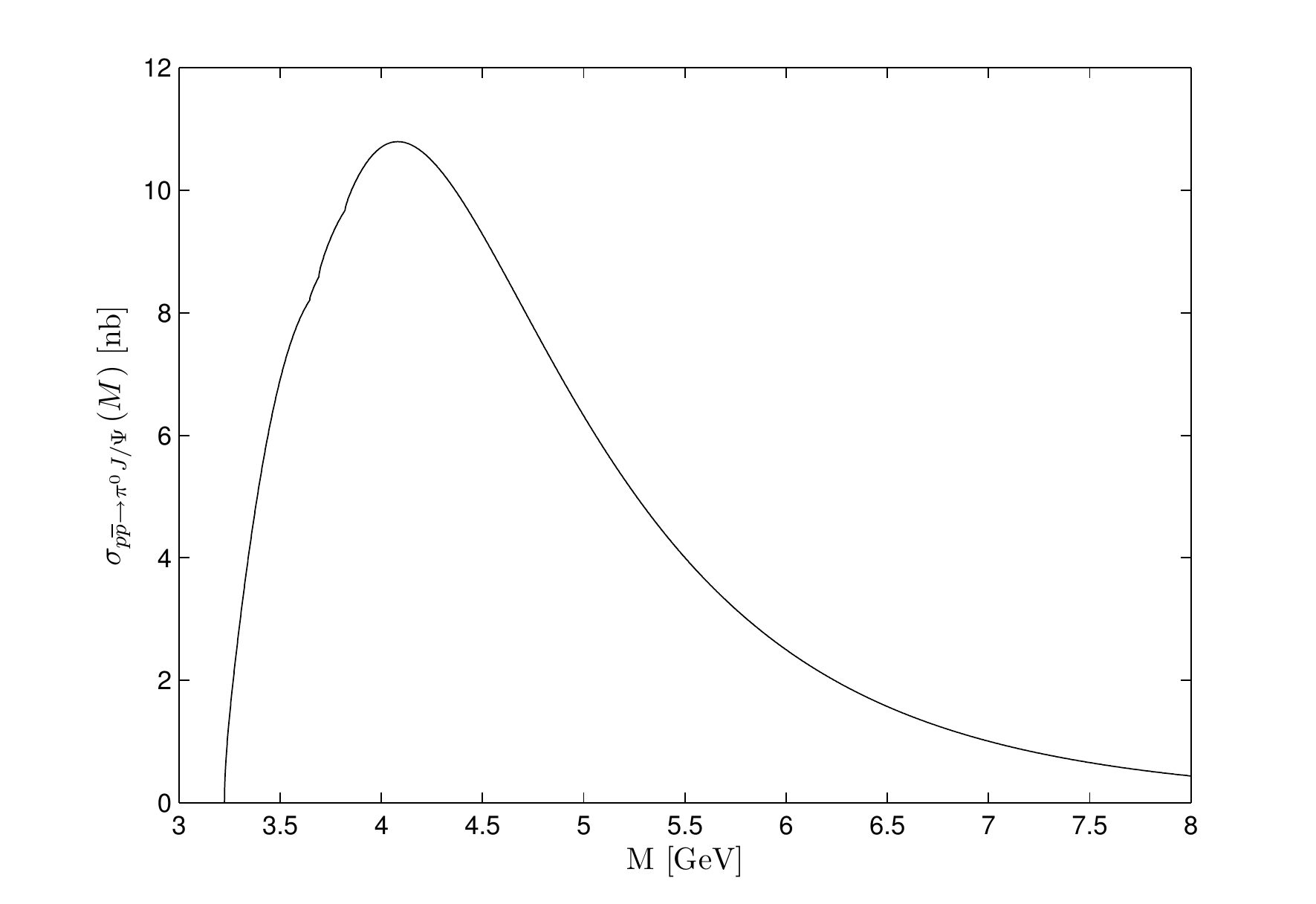}
    \caption{$M$ dependence of the cross section $\sigma_{p\bar p \rightarrow \pi^0 J/\Psi}$}
    \label{Fig:sig_pap_pi0J}  
\end{figure}

As an application we calculated the cross section of the reaction $p +
\bar p \to \pi^0 + J/\Psi$, which is an important ingredient to
describe antiproton induced $J/\Psi$ production -- an important part of
the PANDA/FAIR research plan. For this a reference channel was
needed, where the cross section in the desired energy range is
known. Consequently, we choose the 
$p + \bar p \to n + \bar n$ reaction, for which the cross
section is well known in a relatively wide energy range. The desired
cross section can be expressed as,
\begin{equation}
\sigma_{p\bar p \to \pi^0 J/\Psi}= \frac{\sigma_{p\bar p
    \to n\bar n}}{R_{\pi^0 J/\Psi}^{n \bar n}},
\end{equation}
where $(R_{\pi^0 J/\Psi}^{n \bar n}=W_{n\bar n}/W_{\pi^0 J/\Psi})$ can
be analytically calculated from our model resulting in a similar
expression as in \eref{Eq:R_pp_J}. By using the previously fitted
values for the $P_u,P_d,P_s,P_c$ probabilities given in
\eref{Eq:Pudsc_values}, the resulting cross section can be seen in
\figref{Fig:sig_pap_pi0J}.

\subsection{Proton\,-\,antiproton annihilation at rest}
\label{Subsec:p_antip}

\begin{figure}[!t]
    \includegraphics[width=0.45\textwidth]{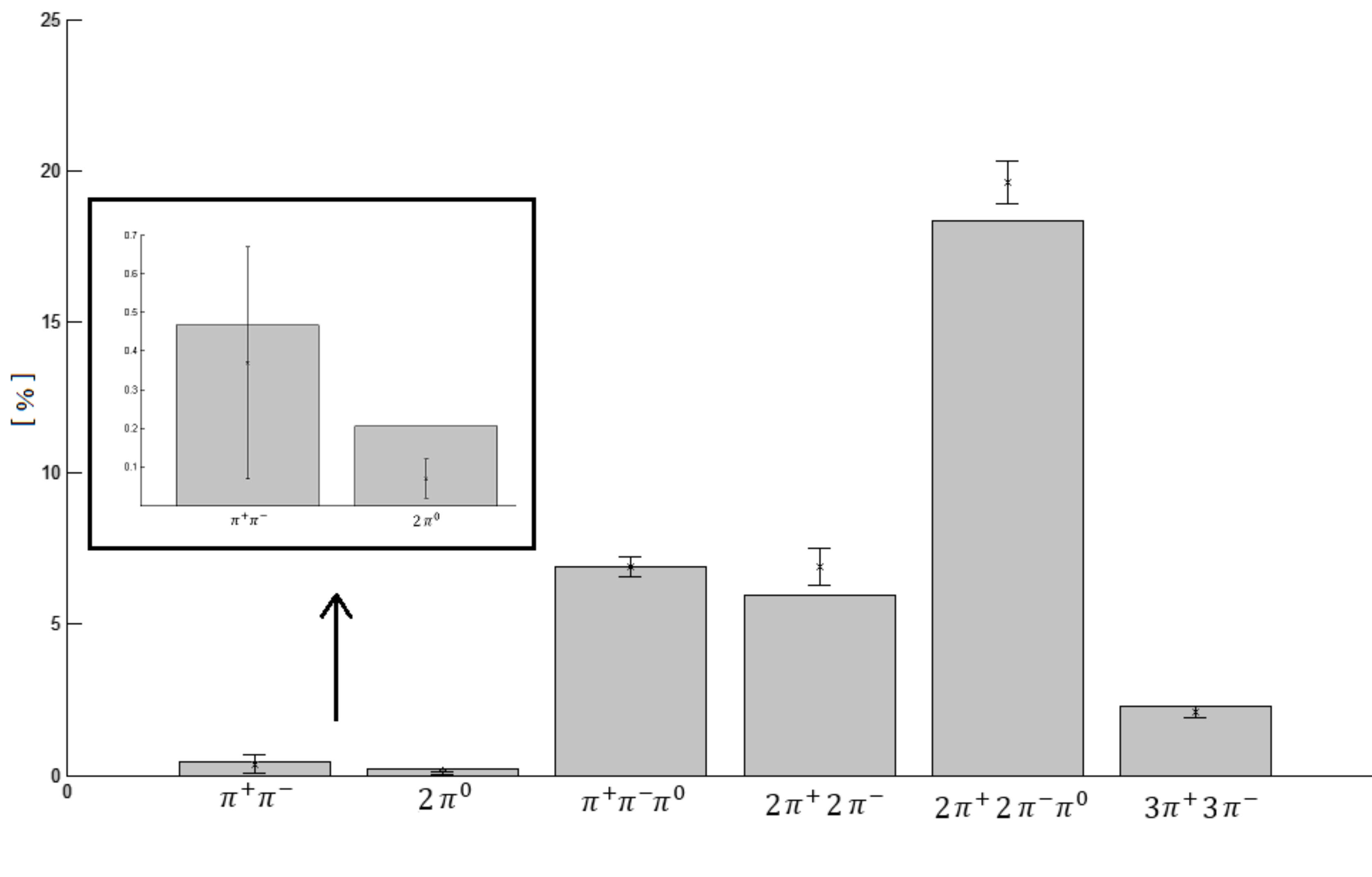}
    \caption{Probabilities of the multiple pion final states of the
      $p\bar p$ collision at rest}
    \label{Fig:pap_multipion}  
\end{figure}

\begin{figure}[!t]
    \includegraphics[width=0.45\textwidth]{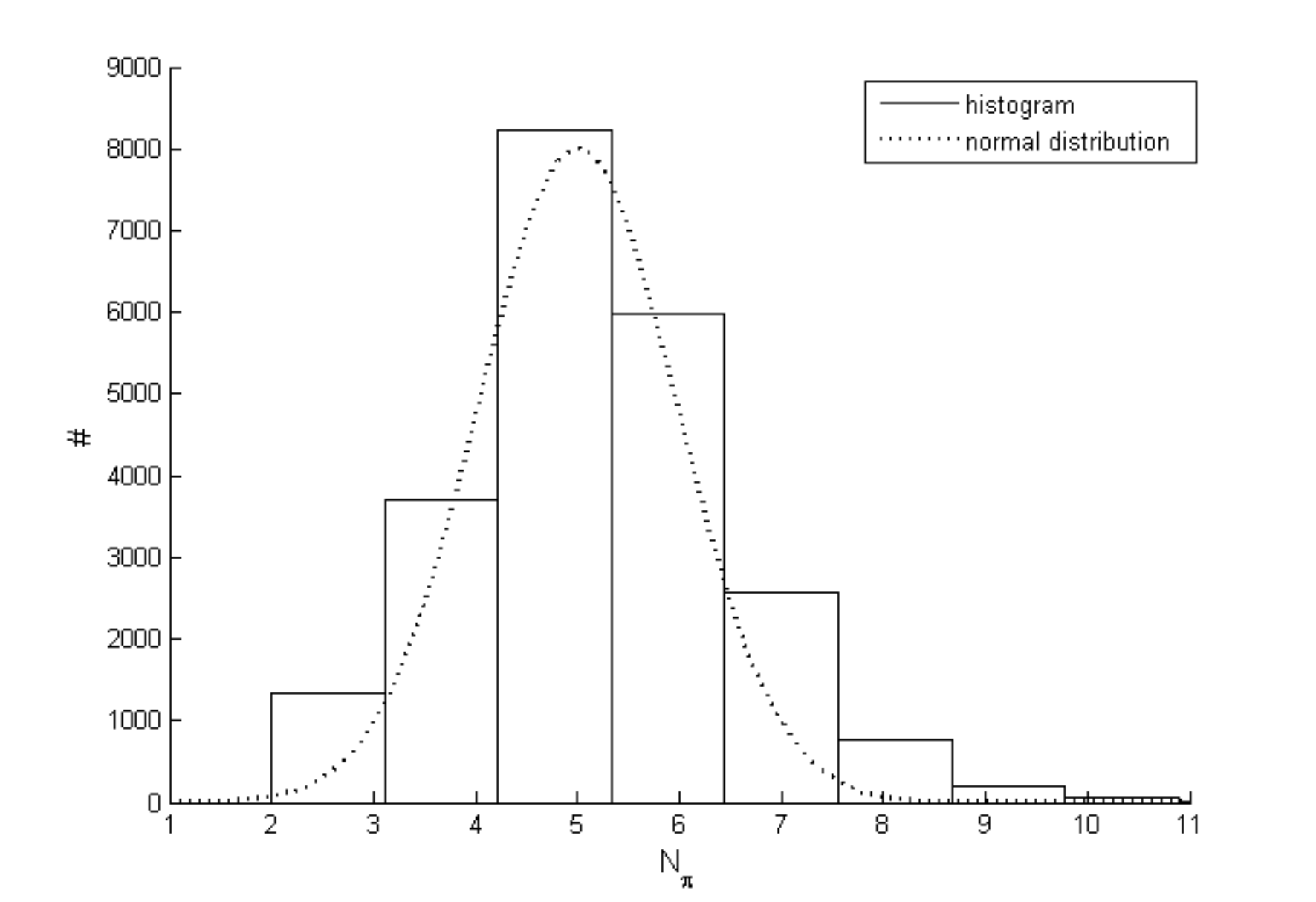}
    \caption{Pion multiplicities in $p\bar p$ annihilation at rest}
    \label{Fig:pion_distr}  
\end{figure}

An important application, and also a good validity check of the model
if we calculate a few, more complicated final states for $p\bar p$
annihilation at rest -- {\it i.e.} at $\sqrt{s} = 2m_p$ -- and compare
with existing measured data; probabilities of the most important
channels and the end state pion distribution can be found in
\cite{Lu:1995,Blumel:1994}. Because of the higher multiplicities --
two to six --, there will be more than one fireball. Owing to the
complexity of the problem the calculations were carried out with the
help of a Monte\,-\,Carlo simulation.
The results are shown in \figref{Fig:pap_multipion}, where the
histograms show the calculated, while the crosses with errorbars the
measured values.
The plot shows a very good agreement, which is promising considering
future applications. The final state pion distribution was also
checked, where again a good match was found to the experimentally
known normal distribution. \figref{Fig:pion_distr} shows the result,
where the following fit could be used,
\begin{eqnarray}
  P(N_{\pi}) \approx
  \frac{1}{\sqrt{2\pi}D}\exp\left(-\frac{\left(N_{\pi}-\left<
          N_{\pi}\right> \right)}{2 D^2} \right),
\end{eqnarray}
where $\left< N_{\pi}\right> \approx 5$ is the average pion
multiplicity, and $D=0.97$ is the standard deviation, which values are
taken from \cite{Dover:1992}.
The main consequence is that the many fireball model seems to be a
good phenomenological approximation for processes with more than three
particle final states, and can be used to estimate various yet
experimentally unknown cross sections.

\section{Conclusions}
\label{Sec:concl}

In this article we proposed a statistical model based on earlier works
\cite{Hagedorn:1965,Hagedorn:1968,Frautschi:1971} and extended by
extra dynamical factors that can be used to describe medium energy
hadronic processes. The original Bootstrap idea of Hagedorn and
Frautschi was extended with quark combinatorics and a specific
many\,-\,fireball decay scheme. The method can be used to calculate
ratios of different processes, and consequently let us -- by using a
known reference channel -- calculate yet unknown cross sections. The
most important parameters of the model are the interaction temperature
$T_0$ and the quark creation probabilities $P_u,P_d,P_s,P_c$, which
were determined by fitting to existing experimental data. Another
parameter, which appears in ratios with different number of fireballs
is the interaction volume $V$ ($=4\pi/(3m_{\pi}^{3})$). To
test our method we calculated some known cross section ratios, which
gave very good match to the measured data. In case of
proton\,-\,antiproton annihilation, where analytical calculations were
not feasible we used a Monte\,-\,Carlo simulations, which also showed
a good agreement to the existing data. This somewhat confirms our idea
of the many\,-\,fireball decay scheme. In the future, we intend to use
this method in a BUU transport code to include unknown elementary
cross sections, especially concentrating on the charmonium formation
processes. The method could also be important to describe possible
many\,-\,body collisions in strongly interacting dense matter.


\section*{Acknowledgments}

The authors were supported by the Hungarian Research Fund (OTKA) under
Contract No. K109462 and by the HIC for FAIR Guest Funds of the Goethe
University Frankfurt. P. K. also acknowledges support from the 
ExtreMe Matter Institute EMMI at the GSI Helmholtzzentrum f\"ur 
Schwerionenforschung, Darmstadt, Germany.

This work is devoted to the memory of Walter Greiner.


\appendix

\section{Explicit form of the phase space integrals}
\label{App:phi23}

\begin{equation}
  \begin{split}
    \Phi_2(&M,m_1,m_2)= V\int d^3 q_1 d^3 q_2
    \delta(M-E_1-E_2)\\
    & \times \delta^{(3)}(q_1+q_2)= \frac{V \pi}{2M^4} \Big(
    M^4-(m_1^2-m_2^2)^2\Big) \\
    & \times \sqrt{\lambda(M^2,m_1^2,m_2^2)} ,
  \end{split}
\end{equation}
where $\lambda(x,y,z)=x^2+y^2+z^2-2xy-2xz-2yz$ is the K\"all\'en function,
and $V$ is the interaction volume which is set to $V=\frac{4\pi}{3 m_{\pi}^{3}}$
\begin{equation}
  \begin{split}
    &\Phi_3(M,m_1,m_2,m_3)=V^2\int d^3q_1 d^3q_2 d^3q_3  \\
    & \times \delta(M-E_1-E_2-E_3) \delta^{(3)}(q_1+q_2+q_3) \\
    & = V^2\left\{\frac{M^5}{120}-\frac{M^3}{12}\sum_{i=1}^3{m_i^2}+{\frac{M^2}{6}}\sum_{i=1}^3{m_i^3}
    \right. \\
    & \left.+{\frac{M}{4}} \Biggl[ \mathop{\sum_{i=1}^3\sum_{j=2}^3}_{i\ne j} m_i^2m_j^2  
        - {\frac{1}{2}}\sum_{i=1}^3{m_i^4} \Biggr]
      +{\frac{1}{30}}\sum_{i=1}^3{m_i^5} \right.  \\
    & \left. -\frac{1}{6}\mathop{\sum_{i=1}^3\sum_{j=1}^3}_{i\ne j} m_i^3 m_j^2\right\}
  \end{split} 
\end{equation}


\section{Formation probability of three fireballs}
\label{App:fr3}

\begin{figure*}[!t]
    \centerline{\includegraphics[width=0.85\textwidth]{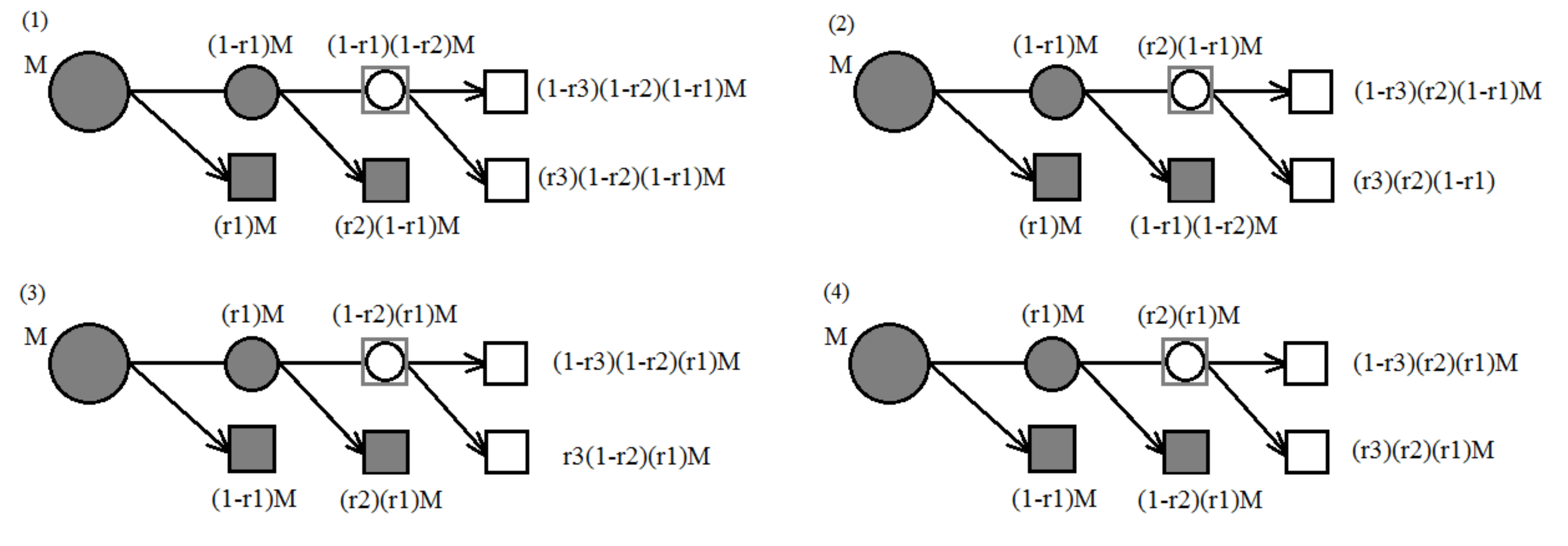}}
    \caption{The decay scheme for the three fireballs formation. The
      four possible subcases are labeled from $(1)$ to $(4)$.}
    \label{Fig:4fireball}  
\end{figure*}

To calculate the three fireball case there are four different
topologies we have to consider, which can be seen in
\figref{Fig:4fireball}.
For the first two decay schemes we can define a common event $A$ as
\begin{equation}
  \begin{split}
    A:\hspace*{1.2cm} \left(r_1>\frac{m_c}{M}\right) \; &\land \; \left(r_1<1-\frac{m_c}{M}\right)  \\
     \land \left(r_2>\frac{m_c}{M(1-r_1)}\right)\; &\land \; \left(r_2<1-\frac{m_c}{M(1-r_1)}\right)
  \end{split}
\end{equation}
This event describes the first part of the decay schemes, which is the
same in the first two topologies. For the first decay scheme have to
define the following set of events.
\begin{eqnarray}
  A1&:&r_3<\frac{m_c}{(1-r_1)(1-r_2)M}, \\
  A2&:&r_3>1-\frac{m_c}{(1-r_1)(1-r_2)M}.
\end{eqnarray}
For the second decay scheme the corresponding events are 
\begin{eqnarray}
  B1&:&r_3<\frac{m_c}{(1-r_1)r_2M}, \\
  B2&:&r_3>1-\frac{m_c}{(1-r_1)r_2M}.
\end{eqnarray} 
The third and the fourth topologies also have a common event, because
they differ only at the end of the chain,
\begin{equation}
  \begin{split}
    C:\hspace*{0.3cm}\left(r_1>\frac{m_c}{M}\right) \; &\land \; \left(r_1<1-\frac{m_c}{M}\right) \\
     \land \left( r_2>\frac{m_c}{Mr_1}\right) \; &\land \; \left( r_2<1-\frac{m_c}{Mr_1} \right).
  \end{split}
\end{equation}
For the third decay scheme the followings events were defined,
\begin{eqnarray}
  C1 &:& r_3 < \frac{m_c}{r_1(1-r_2)M}, \\
  C2 &:& r_3 > 1-\frac{m_c}{r_1(1-r_2)M}.
\end{eqnarray}
The last decay scheme also needs two events to be defined,
\begin{eqnarray}
  D1 &:& r_3 < \frac{m_c}{r_1r_2M}, \\
  D2 &:& r_3 > 1-\frac{m_c}{r_1r_2M}.
\end{eqnarray}
From the previously defined events we can express the total
probability of three\,-\,fireballs formation as,
\begin{eqnarray}
  \label{EqP3}
  \hspace*{-0.3cm}P_3(M)&=&\frac{1}{4} \Big[ P(A1|A)+P(A2|A)-P(A1 \land A2|A) \Big. \nonumber \\
    \Big. &+& P(B1|A)+P(B2|A)-P(B1 \land B2|A) \Big. \nonumber \\
    \Big. &+& P(C1|C)+P(C2|C)-P(C1 \land C2|C) \Big.  \nonumber\\
    \Big. &+& P(D1|C)+P(D2|C)-P(D1 \land D2|C)
    \Big],
\end{eqnarray}
where the $1/4$ factor reflects the four existing subcases. As it can
be seen we have to calculate conditional probabilities for each decay
scheme and then sum up all of the possible events. These probabilities
can be expressed in closed form, however for a few cases numerical
integration is required. The different probabilities are the following
\begin{eqnarray}
  P(B1|A)&=&\left[\int\limits_{\frac{m_c}{M}}^{1-2\frac{m_c}{M}}\int\limits_{\frac{m_c}{M(1-r_1)}}^{1-\frac{m_c}{M(1-r_1)}}dr_1dr_2
    \right.\nonumber \\
    &&\left.\times\left({\frac{m_c}{r_2(1-r_1)M}}\right)\right]_{0<\frac{m_c}{M}<\frac{1}{3}},\\
  P(B2|A)&=& \left[1-\frac{3m_c}{M}+\frac{2m_c}{M}ln\left( {\frac{M}{m_c}-2} \right) \right. \nonumber \\
    &&\left. \hspace*{-2.0cm}-\int\limits_{\frac{m_c}{M}}^{1-2\frac{m_c}{M}}\int
    \limits_{\frac{m_c}{M(1-r_1)}}^{1-\frac{m_c}{M(1-r_1)}}
      dr_1dr_2\left( \frac{m_c}{M(1-r_1)r_2} \right) \right]_{0<\frac{m_c}{M}<\frac{1}{3}},\\
  P(B1 \land B2|A) &=& \nonumber \\
  && \hspace*{-2.0cm}=\left[ \int\limits_{1-3\frac{m_c}{M}}^{1-\frac{m_c}{M}} \int\limits_{\frac{m_c}{M(1-r_1)}}^{1-\frac{m_c}{M(1-r_1)}} dr_1dr_2
    \left( \frac{2m_c}{Mr_2(1-r_1)}-1 \right) \right]_{0<\frac{m_c}{M}<\frac{1}{4}} \nonumber \\
  && \hspace*{-2.0cm}+\left[ \int\limits_{\frac{m_c}{M}}^{1-3\frac{m_c}{M}}
    \int\limits_{\frac{m_c}{M(1-r_1)}}^{\frac{2m_c}{M(1-r_1)}}  dr_1dr_2\left(
    \frac{2m_c}{Mr_2(1-r_1)}-1 \right) \right]_{0<\frac{m_c}{M}<\frac{1}{4}} \nonumber \\
  && \hspace*{-2.0cm}+\left[ \int\limits_{\frac{m_c}{M}}^{1-\frac{m_c}{M}}
    \int\limits_{\frac{m_c}{M(1-r_1)}}^{1-\frac{m_c}{M(1-r_1)}}dr_1dr_2\left(
    \frac{2m_c}{Mr_2(1-r_1)}-1 \right)\right]_{\frac{1}{4}<\frac{m_c}{M}<\frac{1}{2}}\hspace{-0.6cm},
\end{eqnarray}
\begin{eqnarray}
  P(C1|C)&=&
  \left[ \int\limits_{2\frac{m_c}{M}}^{1-\frac{m_c}{M}}\int\limits_{\frac{m_c}{Mr_1}}^{1-\frac{m_c}{Mr_1}}dr_1dr_2
    \times \left( \frac{m_c}{Mr_1(1-r_2)}\right) \right]_{0<\frac{m_c}{M}<\frac{1}{3}}\hspace{-0.6cm},\hspace{0.6cm}
\end{eqnarray}
\begin{eqnarray}
  \hspace{-0.8cm} P(C2|C)&=&\left[
    1-\frac{3m_c}{M}-\frac{2m_c}{M}\ln\left( \frac{M}{2m_c}-\frac{1}{2}
    \right) - \right.  \nonumber \\
    && \hspace{-0.8cm}\left. -\int\limits_{2\frac{m_c}{M}}^{1-\frac{m_c}{M}}\int\limits_{\frac{m_c}{Mr_1}}^{1-\frac{m_c}{Mr_1}}dr_1dr_2\left( 1-\frac{m_c}{Mr_1(1-r_2)}\right) 
    \right]_{0<\frac{m_c}{M}<\frac{1}{3}}\hspace{-1cm},
\end{eqnarray}
\begin{eqnarray}
  P(C1 \land C2|C)&=& \nonumber \\
  &&\hspace*{-2.5cm}=\left[ \int\limits_{\frac{m_c}{M}}^{3\frac{m_c}{M}} \int\limits_{\frac{m_c}{Mr_1}}^{1-\frac{m_c}{Mr_1}}
    dr_1dr_2\left( \frac{2m_c}{Mr_1(1-r_2)}-1 \right)
  \right]_{0<\frac{m_c}{M}<\frac{1}{4}} \nonumber \\
  &&\hspace*{-2.5cm}+\left[ \int\limits_{3\frac{m_c}{M}}^{1-\frac{m_c}{M}}
    \int\limits_{1-\frac{2m_c}{Mr_1}}^{1-\frac{m_c}{Mr_1}}dr_1dr_2\left(
      \frac{2m_c}{Mr_1(1-r_2)}-1
    \right)\right]_{0<\frac{m_c}{M}<\frac{1}{4}} \nonumber \\
  &&\hspace*{-2.5cm}+\left[ \int\limits_{\frac{m_c}{M}}^{1-\frac{m_c}{M}}
    \int\limits_{\frac{m_c}{Mr_1}}^{1-\frac{m_c}{Mr_1}}dr_1dr_2\left(
      \frac{2m_c}{Mr_1(1-r_2)}-1 \right)\right]_{\frac{1}{4}<\frac{m_c}{M}<\frac{1}{2}}\hspace*{-1cm},
\end{eqnarray}
\begin{eqnarray}
  P(D1|C)&=&\left[ \int\limits_{2\frac{m_c}{M}}^{1-\frac{m_c}{M}}\int\limits_{\frac{m_c}{Mr_1}}^{1-\frac{m_c}{Mr_1}}dr_1dr_2
    \left( \frac{m_c}{Mr_1r_2} \right) \right]_{0<\frac{m_c}{M}<\frac{1}{3}}\hspace*{-1cm},\\
  P(D2|C)&=&\left[
    1-\frac{3m_c}{M}-\frac{2m_c}{M}\ln\left( \frac{M}{2m_c}-\frac{1}{2}
    \right) - \right. \nonumber \\
    && \left.\hspace*{-1.5cm}-\int\limits_{2\frac{m_c}{M}}^{1-\frac{m_c}{M}}\int\limits_{\frac{m_c}{Mr_1}}^{1-\frac{m_c}{Mr_1}}dr_1dr_2\left( 1-\frac{m_c}{Mr_1r_2}\right) 
    \right]_{0<\frac{m_c}{M}<\frac{1}{3}},
\end{eqnarray}
\begin{eqnarray}
  P(D1 \land D2|C) &=& \nonumber \\
  && \hspace*{-2.0cm}=\left[ \int\limits_{3\frac{m_c}{M}}^{1-\frac{m_c}{M}}
    \int\limits_{\frac{m_c}{Mr_1}}^{\frac{2m_c}{Mr_1}} dr_1dr_2\left(
      \frac{2m_c}{Mr_1r_2}-1 \right) \right]_{0<\frac{m_c}{M}<\frac{1}{4}} + \nonumber \\
  && \hspace*{-2.0cm}+\left[ \int\limits_{\frac{m_c}{M}}^{3\frac{m_c}{M}}
    \int\limits_{\frac{m_c}{Mr_1}}^{1-\frac{m_c}{Mr_1}}dr_1dr_2\left(
      \frac{2m_c}{Mr_1r_2}-1 \right)\right]_{0<\frac{m_c}{M}<\frac{1}{4}}\nonumber \\
  && \hspace*{-2.0cm}+\left[ \int\limits_{\frac{m_c}{M}}^{1-\frac{m_c}{M}}
    \int\limits_{\frac{m_c}{Mr_1}}^{1-\frac{m_c}{Mr_1}}dr_1dr_2\left(
      \frac{2m_c}{Mr_1r_2}-1
    \right)\right]_{\frac{1}{4}<\frac{m_c}{M}<\frac{1}{2}}\hspace*{-0.5cm}.
\end{eqnarray}


\end{document}